\documentclass[aps,nofootinbib,showpacs,preprintnumbers,amsmath,amssymb]{revtex4}
\usepackage{epsfig}

\begin{document}
\preprint{USM-TH-148, hep-ph/0311202 (v2)}

\title{Calculations of binding energies and masses of heavy quarkonia using renormalon cancellation}

\author{Carlos Contreras}
  \email{carlos.contreras@usm.cl}
\author{Gorazd Cveti\v{c}}
  \email{gorazd.cvetic@usm.cl}
\author{Patricio Gaete}
  \email{patricio.gaete@usm.cl}
\affiliation{Dept.~of Physics, Universidad T\'ecnica
Federico Santa Mar\'{\i}a, Valpara\'{\i}so, Chile}

\date{\today}

\begin{abstract}
We use various methods of Borel integration to
calculate the binding ground energies and masses
of $b {\bar b}$ and $t {\bar t}$ quarkonia.
The methods take into account the leading infrared renormalon
structure of the (hard$+$)soft part of the binding energies 
$E(s)$, and of the corresponding quark pole masses $m_q$, where 
the contributions of these singularities in
$M(s) = 2 m_q + E(s)$ cancel. Beforehand, we
carry out the separation of the binding energy into
its (hard$+$)soft and ultrasoft parts. The resummation
formalisms are applied to expansions of
$m_q$ and $E(s)$ in terms of quantities which do not
involve renormalon ambiguity, such as ${\overline {\rm MS}}$
mass ${\overline m}_q$ and $\alpha_s(\mu)$.
The renormalization scales $\mu$ are different in
calculations of $m_q$, $E(s)$ and $E(us)$. 
The mass ${\overline m}_b$ is extracted, and the binding 
energies $E_{t \bar t}$ and the peak (resonance) energies
$E_{\rm res.}$ for $t \bar t$ production are obtained.

\vspace{1.cm}

\noindent
This is the version v2 as it will appear in Phys. Rev. D.
The changes in comparison to the previous version:
extended discussion between Eqs.~(25) and (26); the paragraph
between Eqs.~(32) and (33) is new and explains the numerical
dependence of the residue parameter on the factorization scale; 
several new references were added; acknowledgments were modified. 
The numerical results are unchanged.

\end{abstract}

\maketitle

\section{Introduction}
\label{intro}
There has been a significant activity in calculation
of binding energies and masses of heavy quarkonia $q {\bar q}$
in recent years. The calculations, based on perturbative expansions,
are primarily due to the knowledge of up to ${\rm N}^2{\rm LO}$ 
term ($\sim\!\alpha_s^3$) of the static quark-antiquark
potential $V(r)$ \cite{Peter:1996ig,Schroder:1998vy}
and partial knowledge of the ${\rm N}^3{\rm LO}$ term there;
the knowledge of the $1/m_q$ and $1/m_q^2$ correction terms
(\cite{Kniehl:2002br} and references therein)
and the ultrasoft gluon contributions to a corresponding 
effective theory ${\rm N}^3{\rm LO}$ Hamiltonian 
\cite{Brambilla:1999qa,Kniehl:1999ud,Kniehl:2002br};
and the knowledge of the the pole mass $m_q$ up to order
$\sim \alpha_s^3$ \cite{Gray:1990yh,Chetyrkin:1999qi}.
Another impetus in these calculations was given by
the observation of the fact that the contributions of the
leading infrared (IR) renormalon singularities (at $b=1/2$)
of the pole mass $m_q$ and of the static potential $V(r)$
cancel in the sum $2 m_q + V(r)$ 
\cite{Hoang:1998nz,Brambilla:1999xf,Beneke:1998rk}
(analogous cancellations were discovered and used in the
physics of mesons with one heavy quark \cite{Neubert:1994wq}).
Consequently, this cancellation effect must be present also 
in the total quarkonium mass $M = 2 m_q + E_{q \bar q}$ 
\cite{Pineda:2001zq,Lee:2003hh}, or more precisely,
in $M(s) = 2 m_q + E(s)$ 
where $E(s)$ is the hard$+$soft part of the binding energy,
i.e., the part which includes the contributions
of relative quark-antiquark momenta $|k^0|, |{\bf k}| \agt
m_q \alpha_s$, i.e., soft/potential scales (predominant)
and higher hard scales (smaller contributions).
In addition, the binding energy has contribution $E_{q \bar q}(us)$ 
from the ultrasoft momenta regime 
$|k^0|, |{\bf k}| \sim m_q \alpha^2_s$. The ultrasoft
contribution is not related to the $b=1/2$ renormalon
singularity, since this singularity has to do with the
behavior of the theory in the region which includes the hard
($\sim m_q$) and soft/potential ($\sim m_q \alpha_s$) scales.

In this work, we numerically calculate 
the binding ground energies $E_{q \bar q}$ 
(separately the $s$ and the $us$ parts)
and the mass $(2 m_q + E_{q \bar q})$ of the heavy $q {\bar q}$ 
system, by taking into account the leading IR
renormalon structure of $m_q$ and $E_{q \bar q}(s)$,
in the spirit of the works of
Refs.~\cite{Pineda:2001zq,Lee:2003hh}.
We combine some features of these two references:
(a) the mass that we use in the perturbation expansions
is a renormalon-free mass 
\cite{Beneke:1998rk,Pineda:2001zq,Bigi:1994em,Hoang:1998ng,Yakovlev:2000pv} 
which we choose
to be the ${\overline {\rm MS}}$ mass
${\overline m}_q \equiv {\overline m}_q(\mu = {\overline m}_q)$;
(b) Borel integrations \cite{Lee:2003hh} are used to perform
resummations. However, before resummations we perform
separation of the soft/potential ($s$) and ultrasoft ($us$)
part of the binding energies, and apply the renormalon-based
Borel resummation only to the $s$ part. The renormalization
scales used in the Borel resummations are $\mu_h \sim m_q$ (hard scale)
for $2 m_q$, and $m_q \alpha_s \alt \mu_s < m_q$ for $E_{q \bar q}(s)$.
The term corresponding to $E_{q \bar q}(us)$ is evaluated
at $\mu_{us} \sim m_q \alpha^2_s$ whenever perturbatively possible. 
Further, the Borel resummations are performed
in three different ways: (a) using a slightly extended version 
of the full bilocal expansion of the type introduced and used
in Refs.~\cite{Lee:2002sn,Lee:2003hh}; 
(b) using a new ``$\sigma$-regularized'' full
bilocal expansion introduced in the present work; 
(c) using the form of the Borel transform where the leading IR
renormalon structure is a common factor of the transform 
\cite{Caprini:1998wg,Cvetic:2001sn} (we call it $R$-method).
The Borel integrations for both $m_q$ and $E_{q \bar q}(s)$
are performed by the same prescription (generalized principal value PV
\cite{Khuri:bf,Caprini:1998wg,Cvetic:2001sn,Cvetic:2002qf})
so as to ensure the numerical cancellation of the 
renormalon contributions in the sum $2 m_q + E_{q \bar q}(s)$.
Furthermore, we demonstrate numerically that in the latter
sum the residues at the renormalons are really 
consistent with the renormalon cancellation
when a reasonable factorization scale parameter for the $s$-$us$
separation is used, while they become incosistent with
the aforementioned cancellation when no such separation is used.  
The obtained numerical results allow us to extract the
mass ${\overline m}_b$ from the known $\Upsilon(1S)$
mass of the $b {\bar b}$ system and to demonstrate that
the $us$ contribution is the major source of uncertainty.
We present also the numerical results for the
ground state binding energy for the scalar and vector
toponium $t {\bar t}$.

In Sec.~\ref{mass} we recapitulate
the calculation of the pole mass $m_q$ in terms of
${\overline m}_q$ and $\alpha_s(\mu_h)$, and summarize
the bilocal method of Refs.~\cite{Lee:2002sn,Lee:2003hh},
with a slight extension in the
renormalon-part of the Borel transform.
In Sec.~\ref{sep} we perform the separation of the
binding ground energy into the soft/potential ($s$) and
the ultrasoft ($us$) part, and in Sec.~\ref{Eqq} we determine 
the $s$-$us$ factorization scale parameter so that the
renormalon residue reproduced from $E_{q \bar q}(s)$
becomes consistent with the renormalon cancellation condition.
In Sec.~\ref{Eqq} we further apply several methods of
the Borel resummation to calculate $E_{b \bar b}(s)$
and $E_{t \bar t}(s)$: the aforementioned bilocal method,
the new ``$\sigma$-regularized'' bilocal method,
as well as the aforementioned $R$-method, and
always using in the expansions ${\overline m}_q$ mass.
We also estimate there the ultrasoft contributions
to the binding energy.
In Sec.~\ref{summ} we compare the obtained results
with some of the results recently published in the literature
and draw conclusions about the main numerical features of
our resummation procedure.

\section{Pole mass}
\label{mass}

Here we redo the calculation of the pole mass $m_q$
in terms of the ${\overline {\rm MS}}$ renormalon-free mass
${\overline m}_q \equiv {\overline m}_q(\mu = {\overline m}_q)$
and of $\alpha_s(\mu,{\overline {\rm MS}})$, using
elements of the approach of Ref.~\cite{Pineda:2001zq}
and the bilocal expansion method of Refs.~\cite{Lee:2002sn,Lee:2003hh}.
In the Borel integration, we choose the (generalized)
principal value (PV) prescription 
\cite{Khuri:bf,Caprini:1998wg,Cvetic:2001sn,Cvetic:2002qf}.
The ratio $S = (m_q/{\overline m}_q - 1)$ has perturbation expansion
in ${\overline {\rm MS}}$ scheme which is at present known 
to order $\sim\!\alpha_s^3$
(Ref.~\cite{Tarrach:1980up} for $\sim\!\alpha_s$;
\cite{Gray:1990yh} for $\sim\!\alpha_s^2$;
\cite{Chetyrkin:1999qi} for $\sim\!\alpha_s^3$)
\begin{subequations}
\label{Sm}
\begin{eqnarray}
S \equiv \frac{m_q}{{\overline m}_q} - 1 & = & \frac{4}{3} a(\mu)
\left[ 1 + a(\mu) r_1(\mu) + a^2(\mu) r_2(\mu) + {\cal O}(a^3) \right] 
\ ,
\label{Smexp}
\\
r_1(\mu) & = & \kappa_1 + \beta_0 L_m(\mu) 
\ ,
\label{r1} 
\\
r_2(\mu) & = & \kappa_2 + ( 2 \kappa_1 \beta_0 + \beta_1) L_m(\mu) 
+ \beta_0^2 L_m^2(\mu) \ ,
\label{r2}
\\
(4/3) \kappa_1 & = & 6.248 \beta_0 - 3.739 \ ,
\label{k1}
\\
(4/3) \kappa_2 &= &   23.497 \beta_0^2 + 6.248 \beta_1 
+ 1.019 \beta_0 - 29.94 \ ,
\label{k2}
\end{eqnarray}
\end{subequations}
where $L_m = \ln(\mu^2/{\overline m}_q^2)$, while
$\beta_0 = (11 - 2 n_f/3)/4$ and $\beta_1 = (102 - 38 n_f/3)/16$
are the renormalization scheme independent coefficients
with $n_f = n_{\ell}$ being the number of light active
flavors (quarks with masses lighter than $m_q$). 
The natural renormalization scale here is $\mu = \mu_h \sim m_q$
(hard scale).

Therefore, the Borel transform $B_S(b)$ is known to order $\sim\!b^2$
\begin{equation}
B_{S}(b; \mu) = \frac{4}{3} \left[ 1 + \frac{r_1(\mu)}{1! \beta_0} b +
\frac{r_2}{2! \beta_0^2} b^2 + {\cal O}(b^3) \right] \ .
\label{BSm1}
\end{equation}
It has renormalon singularities at 
$b = 1/2, 3/2, 2, \ldots, -1, -2, \ldots$ 
\cite{Bigi:1994em,Beneke:1994sw,Beneke:1999ui}.
The behavior of $B_S$ near the leading IR renormalon
singularity $b=1/2$ is determined by the resulting
renormalon ambiguity of $m_q$ which has to have the dimensions 
of energy and should be renormalization scale and
scheme indepenent -- the only such QCD scale being
$const \times \Lambda_{\rm QCD}$ 
\cite{Beneke:1994rs} (cf.~also \cite{Lee:2002sn}).
This scale is proportional to the Stevenson scale
$\widetilde \Lambda$ \cite{Stevenson:1981vj} (cf.~also
\cite{Cvetic:2000mh}). The latter can be obtained in
terms of the strong coupling parameter
$a(\mu; c_2, c_3, \ldots) = \alpha_s(\mu; c_2, c_3, \ldots)/\pi$,
where $c_j = \beta_j/\beta_0$ ($j \geq 2$) are the
parameters characterizing the renormalization scheme,
by solving the renormalization group equation (RGE)
 \cite{Stevenson:1981vj}
\begin{eqnarray}
\frac{d a(\mu)}{d \ln \mu^2 } &=& - \beta_0 a^2(\mu)(1 + c_1 a(\mu)
+ c_2 a^2(\mu) + \cdots ) \quad \Rightarrow
\label{RGE}
\\
\ln \left( \frac{{\widetilde \Lambda}^2}{ \mu^2 } \right)
& = & \frac{1}{\beta_0} \int_0^{a(\mu)} dx \
\left[ \frac{1}{x^2(1 + c_1 x + c_2 x^2 + \cdots )} -
\frac{1}{x^2(1 + c_1 x)} \right] - \frac{1}{\beta_0 a(\mu)} +
\frac{c_1}{\beta_0} \ln \left( \frac{ 1 + c_1 a(\mu) }{c_1 a(\mu)}
\right) \ \Rightarrow
\label{solRGE}
\\
{\widetilde \Lambda}  & = & \mu \ 
\exp \left( - \frac{1}{2 \beta_0 a(\mu)} \right) 
\left( \frac{1 + c_1 a(\mu)}{ c_1 a(\mu)} \right)^{\nu}
\exp\left[ -  \frac{1}{2 \beta_0} \int_0^{a(\mu)} dx \
\frac{( c_2 + c_3 x + c_4 x^2 + \cdots )}{
( 1 + c_1 x ) ( 1 + c_1 x + c_2 x^2 + \cdots )} \right] \ ,
\label{tL1}
\end{eqnarray}
where $\nu = c_1/(2 \beta_0) = \beta_1/(2 \beta_0^2)$; the coefficients
$c_j$ ($j \geq 2$) will be taken here in ${\overline {\rm MS}}$
scheme. Expansion of expression (\ref{tL1}) in powers of
$a(\mu)$ then gives
\begin{equation}
{\widetilde \Lambda} = \mu \exp \left( - \frac{1}{2 \beta_0 a(\mu)}
\right) a(\mu)^{- \nu} c_1^{- \nu} \left[
1 + \sum_{k=1}^{\infty} \ {\widetilde r}_k a^k(\mu) \right] \ ,
\label{tL2}
\end{equation}
where
\begin{subequations}
\label{tr}
\begin{eqnarray}
{\widetilde r}_1 & = & \frac{ ( c_1^2 - c_2) }{2 \beta_0} \ ,
\qquad
{\widetilde r}_2 = \frac{1}{8 \beta_0^2} \left[
( c_1^2 - c_2 )^2 - 2 \beta_0 ( c_1^3 - 2 c_1 c_2 + c_3 ) \right] \ ,
\label{tr1r2}
\\
{\widetilde r}_3 & = & \frac{1}{48 \beta_0^3} \left[
( c_1^2 - c_2 )^3 - 6 \beta_0 (c_1^2 - c_2 )(c_1^3 - 2 c_1 c_2 + c_3 )
+ 8 \beta_0^2 ( c_1^4 - 3 c_1^2 c_2 + c_2^2 + 2 c_1 c_3 - c_4 )
\right] \ .
\label{tr3}
\end{eqnarray}
\end{subequations}
On the other hand, for the uncertainty in $m_q$ from the
$b=1/2$ renormalon singularity to be proportional to
the quantity (\ref{tL2}), this implies that the singular
part of the Borel transform $B_S(b)$ around $b=1/2$
must have the form\footnote{
See, for example, Ref.~\cite{Cvetic:2002qf} for some algebraic details
of obtaining the typical renormalon ambiguity
${\rm Im} S( z = 2 \beta_0 a(\mu) \pm i \varepsilon )$.}
\begin{subequations}
\label{BStc}
\begin{eqnarray}
B_S(b; \mu) & = &  N_m \pi  \frac{\mu}{ {\overline m}_q }
 \frac{1}{ ( 1 - 2 b)^{1 + \nu} } \left[ 1 +
\sum_{k=1}^{\infty} {\widetilde c}_k ( 1 - 2 b)^k \right]
+ B_{S}^{\rm (an.)}(b; \mu) \ ,
\label{BSrenan}
\\
{\widetilde c}_1 & = & \frac{ {\widetilde r}_1 }{ (2 \beta_0 ) \nu } \ ,
\quad
{\widetilde c}_2 = \frac{ {\widetilde r}_2}{ (2 \beta_0 )^2 \nu (\nu -1) }
\ , \quad
{\widetilde c}_3 = \frac{ {\widetilde r}_3}
{ (2 \beta_0 )^3 \nu (\nu -1)(\nu - 2) } \ ,
\label{tc}
\end{eqnarray}
\end{subequations}
and $B_{S}^{\rm (an.)}(b; \mu)$ is analytic on the disk $|b| < 1$.
The ${\overline {\rm MS}}$ coefficients $c_2$ and $c_3$ are
already known \cite{Tarasov:au,vanRitbergen:1997va}, 
but for $c_4$ we have only 
estimates \cite{Ellis:1997sb,Elias:1998bi} obtained 
by Pad\'e-related methods. Ref.~\cite{Ellis:1997sb}
gives $c_4 \approx 97. (n_f\!=\!4); 86. (n_f\!=\!5)$,
and Ref.~\cite{Elias:1998bi} gives 
$c_4 \approx 40. (n_f\!=\!4); 70. (n_f\!=\!5)$. 
However, the estimate of \cite{Ellis:1997sb} is obtained from a
polynomial in $n_f$ with estimated coefficients, where
large cancellations occur between varios terms.
Therefore, we will take as the central values the
estimates of \cite{Elias:1998bi}, with the edges of the
($\pm$) uncertainties covering the values of \cite{Ellis:1997sb}
\begin{subequations}
\label{c4}
\begin{eqnarray}
c_4 & = & 40 \pm 60 \qquad (n_f\!=\!4) \ ,
\label{c4nf4}
\\
c_4 & = & 70 \pm 20 \qquad (n_f\!=\!5) \ .
\label{c4nf5}
\end{eqnarray}
\end{subequations}
Thus, ${\widetilde c}_3$ can be obtained via Eqs.~(\ref{tr3}), (\ref{tc}):
${\tilde c_3} = 0.01 \pm 0.17 \ (n_f\!=\!4)$; 
$ -0.20 \pm 0.08 \ (n_f\!=\!5)$.
The values of $c_k$'s and ${\widetilde c}_k$'s are given in
Table \ref{table1}.
\begin{table}
\caption{\label{table1}The ${\overline {\rm MS}}$ RGE coefficients
$c_k = \beta_k/\beta_0$ ($k=1,2,3,4$) and renormalon
coefficients $\nu$ and ${\widetilde c}_j$ ($j=1,2,3$)
for the $b \bar b$ ($n_f=4$) and $t \bar t$ ($n_f=5$)
system.}
\begin{ruledtabular}
\begin{tabular}{crrrrrrrr}
 $n_f$ & $c_1$ & $c_2$ & $c_3$ & $c_4$ & $\nu$ &
${\widetilde c}_1$ & ${\widetilde c}_2$ & ${\widetilde c}_3$ \\
\hline
4 & 1.5400 & 3.0476 & 15.0660 & $(40 \pm 60)$ &
0.3696 &- 0.1054 & 0.2736 & 
$(0.01 \mp 0.17)$
\\
5 & 1.2609 & 1.4748 & 9.8349 &  $(70 \pm 20)$ &
0.3289 & 0.0238 & 0.3265 & 
$(-0.20 \mp 0.08)$
\\
\end{tabular}
\end{ruledtabular}
\end{table}
Now, the (full) bilocal method \cite{Lee:2002sn} consists of taking
in the expansion (\ref{BSrenan}) for the analytic part
$B_{S}^{\rm (an.)}$ a polynomial in powers of $b$, so that
the expansion of $B_S$ around $b\!=\!0$ agrees with
expansion (\ref{BSm1}). For that, the residue parameter
$N_m$ in Eq.~(\ref{BSrenan}) has to be determined.
Using the idea of Refs.~\cite{Lee:1996yk}
it was estimated with a high precision in 
Refs.~\cite{Pineda:2001zq,Lee:2003hh,Cvetic:2003wk}:
\begin{equation}
N_m = \frac{{\overline m}_q}{\mu} \frac{1}{\pi} R_S(b=1/2) \ ,
\label{Nmform}
\end{equation}
where, according to (\ref{BSrenan})
\begin{equation}
R_S(b; \mu) \equiv  (1 - 2 b)^{1 + \nu} B_{S}(b; \mu) \ .
\label{RSm}
\end{equation}
In this work, in applications of the bilocal and related methods, we
will use the value of $N_m$ as estimated in Ref.~\cite{Cvetic:2003wk},
which used for $R_S(b)$ truncated perturbation series (TPS)
and Pad\'e approximation $[1/1]$:
\begin{subequations}
\label{Nmnf}
\begin{eqnarray}
N_m(n_f\!=\!4) &=& 0.555 \pm 0.020 \ ,
\label{Nmnf4}
\\
 N_m(n_f\!=\!5) &=& 0.533 \pm 0.020 \ .
\label{Nmnf5}
\end{eqnarray}
\end{subequations} 
The bilocal expansion (\ref{BSrenan}) has then for the
analytic part the polynomial
\begin{subequations}
\label{BSan}
\begin{eqnarray}
B_{S}^{\rm (an.)}(b; \mu) & = & h^{(m)}_0 + \frac{h^{(m)}_1}{1! \beta_0} b
+ \frac{h^{(m)}_2}{2! \beta_0^2} b^2 \ ,
\label{BSanexp}
\\
h^{(m)}_k & = & \frac{4}{3} r_k  - \pi N_m \frac{\mu}{{\overline m}_q}
(2 \beta_0)^k \sum_{n=0}^3 {\widetilde c}_n 
\frac{ \Gamma ( \nu + k + 1 - n) }{ \Gamma(\nu + 1 - n) } \ ,
\label{hms}
\end{eqnarray}
\end{subequations}
where, by convention, $r_0 = {\widetilde c}_0 = 1$.
We can then take for $B_S$ the bilocal formula, i.e., Eqs.~(\ref{BSrenan}) 
and (\ref{BSan}) with the expansion around $b\!=\!1/2$ in the singular 
renormalon part truncated with the term ${\widetilde c}_3 ( 1 - 2 b)^3$
\begin{eqnarray}
B_S(b; \mu)^{\rm (biloc.)} & = &  N_m \pi  \frac{\mu}{ {\overline m}_q }
 \frac{1}{ ( 1 - 2 b)^{1 + \nu} } \left[ 1 +
\sum_{k=1}^3{\widetilde c}_k ( 1 - 2 b)^k \right]
+ \sum_{k=0}^2 \ \frac{h_k^{(m)}}{k! \ \beta_0^k} \ b^k \ .
\label{BSbiloc}
\end{eqnarray}
Applying the (generalized) principal value (PV)
prescription for the Borel integration
\begin{equation}
S(b) = \frac{1}{\beta}_0 {\rm Re} 
\int_{\pm i \varepsilon}^{\infty \pm i \varepsilon} \ db 
\ \exp \left( - \frac{b}{\beta_0 a(\mu)} \right) \ B_S(b; \mu) \ ,
\label{BSint}
\end{equation}
we obtain the pole mass $m_q$ in terms of the mass ${\overline m}_q$.
The numerical integration is performed, using the
Cauchy theorem, along a ray
with a nonzero finite angle with respect to the $b > 0$ axis,
in order to avoid the vicinity of the pole 
(as explained, for example, in Refs.~\cite{Cvetic:2001sn}).

\begin{figure}[htb]
\begin{minipage}[b]{.49\linewidth}
 \centering\epsfig{file=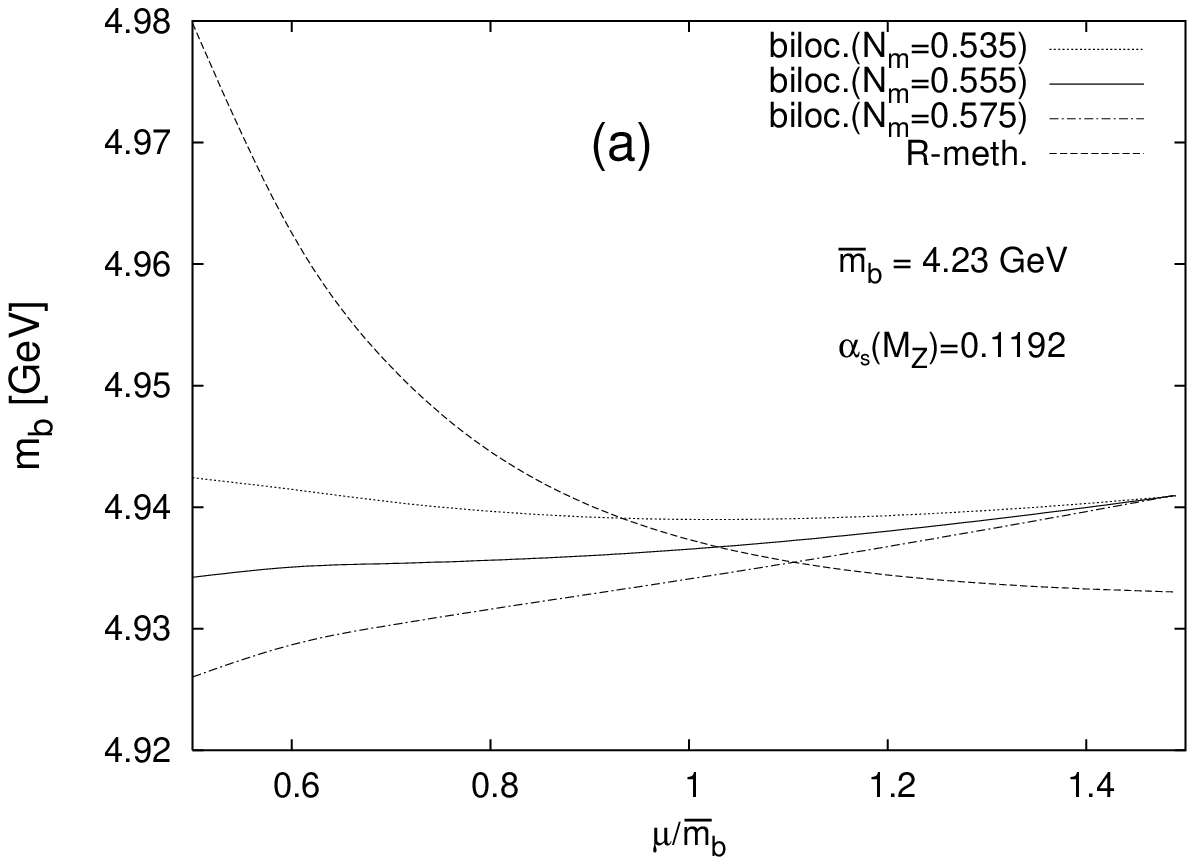,width=\linewidth}
\end{minipage}
\begin{minipage}[b]{.49\linewidth}
 \centering\epsfig{file=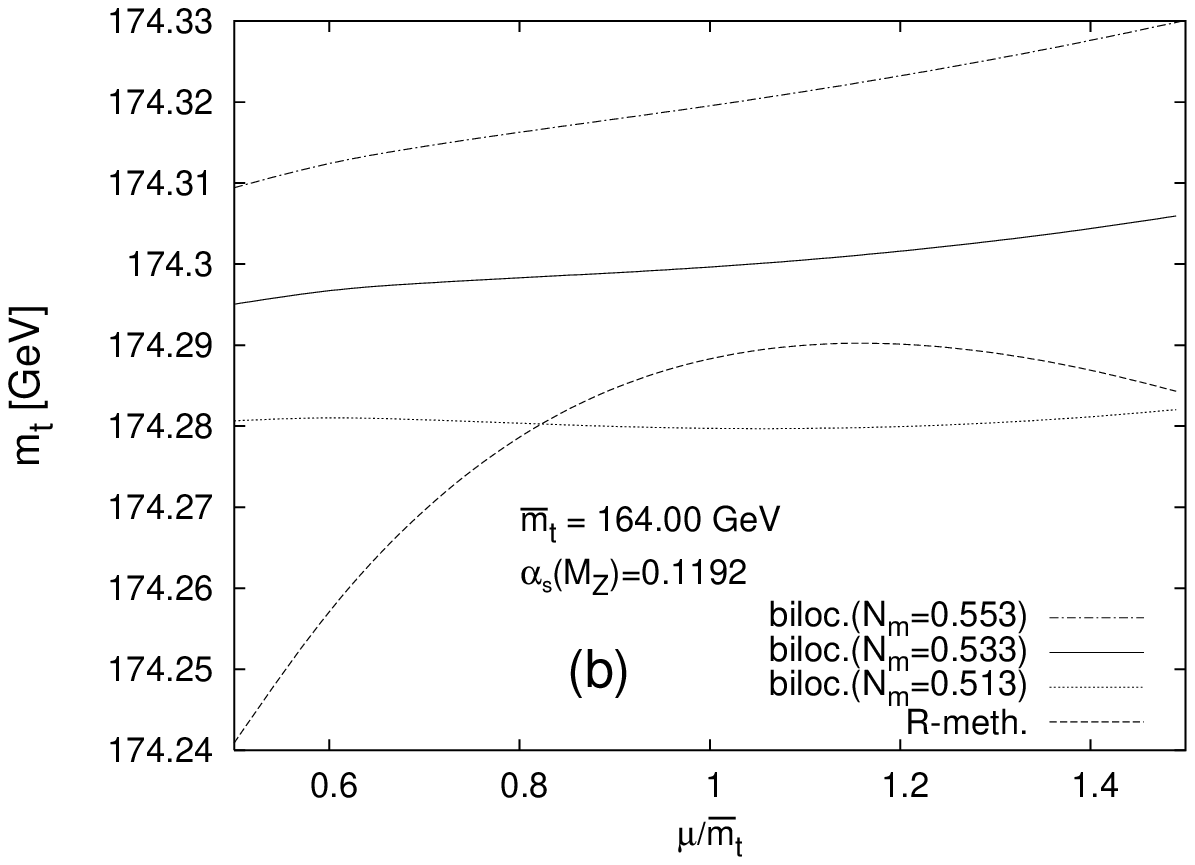,width=\linewidth}
\end{minipage}
\vspace{0.2cm}
\caption{\footnotesize The (PV) pole mass 
of the (a) bottom and (b) top quark, as function of the 
renormalization scale $\mu$. The input parameters
used were ${\overline m}_b = 4.23$ GeV, ${\overline m}_t = 164.00$ GeV,
respectively; the residue parameter values (\ref{Nmnf}) were used
for the bilocal method. The reference value for $\alpha_s$ 
(in ${\overline {\rm MS}}$) was taken to be $\alpha_s(m_{\tau}) = 0.3254$
(\cite{Cvetic:2001ws}) corresponding to $\alpha_s(M_Z) = 0.1192$.}
\label{mqmu.fig}
\end{figure}
In Figs.~\ref{mqmu.fig} (a), (b), we present the resulting (PV) pole
masses of the $b$ and $t$ quarks, as function of the renormalization
scale $\mu$. The spurious $\mu$-dependence is very weak. 
In addition, results of another method (``R''-method )
are presented in Figs.~\ref{mqmu.fig} (a), (b), with the $\mu$-dependence
stronger in the low-$\mu$ region ($\mu/{\overline m}_q < 1$).
The R-method (applied in other contexts in 
Refs.~\cite{Caprini:1998wg,Cvetic:2001sn}) consists in
the Borel integration of the function (\ref{RSm})
\begin{equation}
S = \frac{1}{\beta_0} {\rm Re} 
\int_{\pm i \varepsilon}^{\infty \pm i \varepsilon} \ db 
\ \exp \left( - \frac{b}{\beta_0 a(\mu)} \right)
\ \frac{R_S(b; \mu)}{(1 - 2 b)^{1 + \nu}} \ ,
\label{BRSint}
\end{equation}
where for $R_S(b)$ the corresponding (NNLO) TPS is used.
When we take ${\overline m}_b = 4.23$ GeV and ${\overline m}_t = 164.00$
GeV and we vary the values of the residue parameter $N_m$ according
to Eqs.~(\ref{Nmnf}), the bilocal method gives,
at $\mu/{\overline  m}_q = 1$, variation
$\delta m_b = \mp 3$ MeV and $\delta m_t = \pm 20$ MeV.
When the central values of $N_m$ (\ref{Nmnf}) are used, the
variation of the obtained values of $m_q$ with $\mu$,
when $\mu/{\overline m}_q$ grows from 1.0 to 1.5, is
about $5$ MeV and $6$ MeV for $m_b$ and $m_t$, respectively
(for $R$-method: $4$ MeV and $6$ MeV). When $c_4$ is varied
according to (\ref{c4}), the variation is about $\mp\!2$ and $\mp\!1$ MeV
for $m_b$, $m_t$, respectively. The uncertainty in $\alpha_s$
can be taken as
$\alpha_s(m_{\tau}) = 0.3254 \pm 0.0125$ \cite{Cvetic:2001ws},
corresponding to $\alpha_s(M_Z) = 0.1192 \pm 0.0015$.
This uncertainty is by far the major source in the variation
of the pole masses: ($\delta m_b)_{\alpha_s} = ^{+135}_{-148}$ MeV
for bilocal method ($^{+137}_{-150}$ MeV for $R$-method),
and ($\delta m_t)_{\alpha_s} = ^{+161}_{-171}$ MeV
for bilocal method ($ ^{+161}_{-170}$ MeV for $R$-method).

The natural renormalization scale $\mu$ here is
a hard scale $\mu\!\sim\!{\overline m}_q$, and will be
denoted later in this work as $\mu_m$ in order to distinguish
if from the ``soft'' renormalization scale $\mu$ used in the
analogous renormalon-based resummations of the 
(hard$+$)soft binding energy $E_{q \bar q}(s)$
(${\overline m}_q > \mu \agt {\overline m}_q \alpha_s$)
in Sec.~\ref{Eqq}. The fact that the two renormalization scales
are different does not affect the mechanism of the
($b\!=\!1/2$) renormalon cancellation in the bilocal calculations of
the meson mass $(2 m_q\!+\!E_{q \bar q}(s))$, because the
renormalon ambiguity in each of the two terms is 
renormalization scale independent $\sim\!{\widetilde \Lambda}$,
as seen by Eqs.~(\ref{tL2})--(\ref{BStc}).
On the other hand, if $R$-type methods (\ref{BRSint}) 
[cf.~also Eq.~(\ref{Nmform})] are applied 
for the resummations of $2 m_q$ and $E_{q \bar q}(s)$, the
renormalon ambiguities are renormalization scale independent 
in the approximation of the one-loop RGE running, and the
renormalon cancellation is true at this one--loop level.

\section{Separation of the soft and ultrasoft contributions}
\label{sep}

The perturbation expansion of the (hard $+$ soft $+$ ultrasoft)
binding energy $E_{q \bar q}$ of the $q {\bar q}$ heavy quarkonium
vector ($S=1$) or scalar ($S=0$) ground state ($n=1, \ell=0$)
up to the ${\rm N}^3 {\rm LO}$ ${\cal O}(m_q \alpha_s^5)$
was given in \cite{Penin:2002zv}, where previous
results of Ref.~\cite{Kniehl:2002br} were used. The latter
reference used in part the results of  
Refs.~\cite{Fischler:1977yf,Appelquist:es,Kniehl:2001ju,Schroder:1998vy}
(static potential) and of
Refs.~\cite{Gupta:pd,Titard:1993nn,Czarnecki:1997vz,Melnikov:1998ug,Penin:1998zh}
(binding energy).
Ref.~\cite{Penin:2002zv} (and \cite{Kniehl:2002br})
employed the method of threshold expansion where the
integrations were performed in $(3 - 2 \varepsilon)$ dimensions.
The reference mass scale used was the pole mass $m_q$.
The ground state energy expansion has the form 
\begin{eqnarray}
E_{q \bar q} & = & - \frac{4}{9} m_q \pi^2 a^2(\mu) 
{\Big \{} 1 + a(\mu) \left[ k_{1,0} + k_{1,1} L_p(\mu) \right] +
a^2(\mu) \left[ k_{2,0} + k_{2,1} L_p(\mu) + k_{2,2} L_p^2(\mu) \right]
\nonumber\\
&& + a^3(\mu) \left[ k_{3,0} + k_{3,1} L_p(\mu) + k_{3,2} L_p^2(\mu) 
+ k_{3,3} L_p^3(\mu) \right] + {\cal O}(a^4) {\Big \}} \ ,
\label{Eqqexp}
\end{eqnarray}
where 
\begin{eqnarray}
L_p(\mu) &=& \ln \left( \frac{\mu}{\frac{4}{3} m_q \pi a(\mu) } \right) \ .
\label{Lp}
\end{eqnarray}
The expressions for the coefficients
$k_{i,j}$ of perturbation expansion (\ref{Eqqexp}) for the ground state
binding energy of the quarkonium ($n=1$; $\ell=0$; $S=1$ or $0$)
are given below.
The NLO and NNLO terms were obtained in 
Refs.~\cite{Titard:1993nn,Czarnecki:1997vz,Melnikov:1998ug,Penin:1998zh}.
The ${\rm N}^3 {\rm LO}$ terms were obtained in
Ref.~\cite{Penin:2002zv} --
their Eqs.~(6) and (12), but now written in numerically more explicit
form (and with $N_c=3$).
\begin{subequations}
\label{kij}
\begin{eqnarray}
k_{1,1} & = & 4 \beta_0 \ ,
\qquad
k_{1,0} =  \left( \frac{97}{6} - \frac{11}{9} n_f \right) \ ,
\label{k1s}
\\
k_{2,2} & = & 12 \beta_0^2 \ , 
\qquad 
k_{2,1} = \frac{927}{4} - \frac{193}{6} n_f + n_f^2  \ ,
\label{k2221}
\\
k_{2,0} & = & 361.342 - 40.9649 \ n_f + 1.16286 \ n_f^2  - 11.6973 \ S (S+1) \ ,
\label{k20}
\\
k_{3,3} & = & 32 \beta_0^3 \ ,
\qquad
k_{3,2} = \frac{4521}{2} - \frac{10955}{24} n_f + \frac{1027}{36} n_f^2 -
\frac{5}{9} n_f^3 \ ,
\label{k3332}
\\
k_{3,1} & = & 7242.3 - 1243.95 \ n_f + 69.1066 \ n_f^2  - 1.21714 \ n_f^3 
+ \frac{\pi^2}{2592} ( - 67584 + 4096 \ n_f ) \ S (S+1) \ ,
\label{k31}
\\
k_{3,0} & = & {\bigg [} \left( 7839.82 - 1223.68 \ n_f + 69.4508 \ n_f^2 
- 1.21475 \ n_f^3 \right) 
\nonumber\\
&& + (- 109.05 + 4.06858 \ n_f) \ S(S+1)
- \frac{\pi^2}{18} 
\left( -1089 + 112 \ S(S+1) \right) \ln \left( a(\mu) \right) 
+ 2 \frac{a_3}{4^3} 
{\bigg ]} \ ,
\label{k30}
\end{eqnarray}
\end{subequations}
Here, $a_3$ is the hitherto unknown three--loop contribution
coefficient to the QCD static potential $V_{q \bar q}(r)$,
whose values have been estimated by various methods in
Refs.~\cite{Chishtie:2001mf,Pineda:2001zq,Lee:2003hh,Cvetic:2003wk}.
We will use in this work the estimates of Ref.~\cite{Cvetic:2003wk},
obtained from the condition of renormalon cancellation
in the sum $(2 m_q + V_{q \bar q}(r))$
\begin{subequations}
\label{a3nf}
\begin{eqnarray}
\frac{1}{4^3} \ a_3(n_f\!=\!4) & \approx & 
86. \pm 23. \ ,
\label{a3nf4}
\\
\frac{1}{4^3} \ a_3(n_f\!=\!5) & = & 
62.5 \pm 20. \ .
\label{a3nf5}
\end{eqnarray}
\end{subequations}
The coefficients (\ref{kij}) in the expansion (\ref{Eqqexp})
originate from quantum effects from various scale regimes 
of the participating particles:
(a) the hard scales ($\sim\!m_q$);
(b) the soft and potential scales
where the three momenta are $|{\bf q}|\!\sim\!m_q \alpha_s$
($|q^0|\!\sim\!m_q \alpha_s$ in the soft
and $|q^0|\!\sim\!m_q \alpha_s^2$ in the potential regime);
(c) ultrasoft scales where
$|q^0|$ and $|{\bf q}|$ are both $\sim\!m_q \alpha_s^2$.
The coefficients are dominated by the soft scales;
the hard scales start contributing at the NNLO \cite{Kniehl:2002br}
and are numerically smaller. For this reason, in this
work we will usually refer to the combined soft and hard
regime contributions to the binding energy as simply
soft ($s$) contribution $E_{q\bar q}(s)$.
Strictly speaking, it is only the pure soft regime that
contributes to the $b=1/2$ renormalon.
However, for simplicity,
in our renormalon-based resummations we will
resum the hard$+$soft contributions $E_{q \bar q}(s)$
together, not separately. This will pose no problem,
since the hard regime, being clearly perturbative, 
is not expected to deteriorate the convergence properties
of the series for $E_{q \bar q}(s)$. The natural
renormalization scale $\mu$ in the resummations of $E_{q \bar q}(s)$
is expected to be closer to the soft scale
($m_q \alpha_s \alt \mu < m_q$). 

On the other hand, the ${\rm N}^3 {\rm LO}$ coefficient
$k_{3,0}$ obtains additional contributions from the 
from the ultrasoft ($us$) regime.
The leading ultrasoft contribution comes from the exchange of 
an ultrasoft gluon in the heavy quarkonium 
\cite{Kniehl:1999ud,Brambilla:1999xf}. It consists of two parts:
\begin{enumerate}
\item
The retarded part, which cannot be interpreted in terms of
an instantaneous interaction
\begin{equation}
\frac{1}{\pi^3} k_{3,0}(us,{\rm ret.}) =  
- \frac{2}{3 \pi} \left( \frac{4}{3} \right)^2 L_1^E \approx + 41.014 \ , 
\label{k30usr}
\end{equation}
where $L_1^E \approx - 81.538$ is the QCD Bethe logarithm 
-  see Refs.~\cite{Kniehl:1999ud,Kniehl:2002br}.
\item
The non-retarded part can be calculated as expectation value
of the $us$ effective Hamiltonian ${\cal H}^{us}$ in the
Coulomb (i.e., leading order) ground state $|1 \rangle$, 
where ${\cal H}^{us}$ (in momentum space) was derived in 
Refs.~\cite{Kniehl:1999ud,Kniehl:2002br}. Direct calculation of the
expectation value, here in coordinate space, then gives:
\begin{subequations}
\label{k30usnr}
\begin{eqnarray}
\lefteqn{
\frac{1}{\pi^3} k_{3,0}(us,{\rm nonret.}) =
- \frac{9}{4 \pi^5}  \frac{1}{m_q a^5(\mu)}
\langle 1 | {\cal H}^{us} | 1 \rangle =
\frac{2}{ \pi^5 m_q a^4(\mu)} \left[ 
\frac{1}{2} \ln \frac{\mu_f^2}{(E_1^C)^2}
+ \frac{5}{6} - \ln 2 \right]
}
\nonumber\\
&& \times 
{\Big \{} 
- \frac{27 \pi^3}{8} a^3(\mu) \langle 1 | \frac{1}{r} | 1 \rangle 
- 17 \pi^2 \frac{a^2(\mu)}{m_q} \langle 1 | \frac{1}{r^2} | 1 \rangle
+ \frac{4 \pi^2}{3} \frac{a(\mu)}{m_q^2}
\langle 1 | \delta({\bf r}) | 1 \rangle + 
3 \pi \frac{a(\mu)}{m_q^2} \langle 1 | \{ \Delta_{\bf r}, \frac{1}{r}
\} | 1 \rangle {\Big \}} \ ,
\label{k30usnr1}
\\
& = & - 14.196 \left[ \ln \left( \frac{ \mu_f}{m_q \alpha_s^2(\mu)}
\right) + 0.9511 \right] \ .
\label{k30usnr2}
\end{eqnarray} 
\end{subequations}
Here, $E_1^C = -(4/9) m_q \alpha_s^2(\mu)$ is the Coulomb
energy of the state $|1 \rangle$, and $\mu_f$ is the factorization
energy between the soft ($\!\sim m_q \alpha_s$) and ultrasoft
($\!\sim m_q \alpha^2_s$) scale.
\end{enumerate}

In Ref.~\cite{Kniehl:2002br}, the authors included in the
ultrasoft part of the Hamiltonian additional terms
$\delta  {\cal H}^{us}$ which contained contributions from
the soft regime. These terms arised because of
their use of a method called threshold expansion 
\cite{Beneke:1997zp} 
where the integrations over potential momenta are 
not performed in three dimensions but in
$(d - 1 ) = (3 - 2 \varepsilon)$ dimensions. However,
their method gave in the soft regime also
the same additional terms, but with negative sign 
(including logarithmic terms not associated with
IR-divergent integrals -- unphysical). Since they
were interested in the total sum of contributions
from various regimes, the method gave the correct
result, as emphasized by the authors there.

The $s$--$us$ factorization scale $\mu_f$ can be
estimated as being roughly in the middle between
the $s$ and $us$ energies on the logarithmic scale
\cite{Cvetic:2003wk}
\begin{equation}
\mu_f \left[ \approx (E_{\rm S} E_{\rm US})^{1/2} \right] = \kappa \ m_q 
\alpha_s(\mu_s)^{3/2} \ ,
\label{muf}
\end{equation}
where $\kappa \sim 1$ and $\mu_s \approx E_{\rm S}$ ($\alt\!\mu$).
Therefore, the ultrasoft part of the ${\rm N}^3 {\rm LO}$
coefficient $k_{3,0}$ can be rewritten, by Eqs.~(\ref{k30usr}),
(\ref{k30usnr}) and (\ref{muf}), in terms of the
$s$--$us$ parameter $\kappa$ as
\begin{equation}
\frac{1}{\pi^3} k_{3,0}(us) = 27.512 + 7.098 \ln (\alpha_s(\mu_s))
- 14.196 \ln ( \kappa ) \ .
\label{k30us}
\end{equation}
The soft scale $\mu_s$ appearing here will be fixed
by the condition $\mu_s = (4/3) {\overline m}_q \alpha_s(\mu_s)$.

The formal perturbation expansions for the separate
soft and ultrasoft parts of the ground state binding energy (\ref{Eqqexp})
are then
\begin{subequations}
\label{Eqqsus}
\begin{eqnarray}
E_{q \bar q}(s) & = & - \frac{4}{9} m_q \pi^2 a^2(\mu)
{\Big \{} 1 + \sum_{i=1}^2 a^i(\mu) \sum_{j=0}^i k_{i,j} L_p(\mu)^j
+ a^3(\mu) \sum_{j=1}^3 k_{3,j} + 
a^3(\mu)  \left[ k_{3,0}\!-\!k_{3,0}(us) \right] + {\cal O}(a^4)
{\Big \}} \ ,
\label{Eqqs}
\\
E_{q \bar q}(us) & = & - \frac{4}{9} m_q \pi^2 a^2(\mu)
{\big \{}  a^3(\mu) k_{3,0}(us)  + {\cal O}(a^4)
{\big \}} \ .
\label{Eqqus}
\end{eqnarray}
\end{subequations}
The energy $E_{q \bar q}(s)$ (\ref{Eqqs})
contains the leading IR renormalon effects,
and $E_{q \bar q}(us)$ (\ref{Eqqus}) does not.
In these expressions, the common factor is the soft scale
$\mu_p(\mu) = (4/3) m_q \alpha_s(\mu)$ which is also present
as the reference scale in the logarithms 
$L_p(\mu) = \ln (\mu/\mu_p(\mu) )$ appearing with the coefficients 
$k_{i,j}$ (when $j \geq 1$) in Eqs.~(\ref{Eqqexp}), (\ref{Lp}).
This soft scale is equal to $2/a_{\rm B.}(\mu)$ where
$a_{\rm B.}$ is the (Bohr) radius of the heavy quarkonium.
The renormalization scale $\mu$ in Eq.~(\ref{Eqqs}) is of the
order of the soft scale or above. 
We will re-express $m_q$ everywhere in $E_{q \bar q}$ with the renormalon-free
mass ${\overline m}_q$ (\ref{Sm}), and will
consider the dimensionless soft-energy quantity 
$E_{q \bar q}(s)/{\overline m}_q$. 

The expansion of $E_{q \bar q}(s)/{\overline m}_q =
\sum_{0}^{\infty} {\tilde r_n} (\mu) a^{n+2}(\mu)$
has at large orders the seemingly peculiar feature of the so-called
''power mismatch'' \cite{Hoang:1998ng} (see also \cite{Kiyo:2000fr}): 
when this sum is added to the expansion (\ref{Sm})
$2 m_q/{\overline m}_q = [ 2 + (8/3)\sum_{0}^{\infty} r_n(\mu) a^{n+1}(\mu)]$ , 
the coefficient ${\tilde r}_n(\mu)$ at powers
$a^{n+2}(\mu)$ of $E_{q \bar q}(s)/{\overline m}_q$ must be 
combined with the coefficient $(8/3) r_n(\mu)$
at powers $a^{n+1}(\mu)$ of $2 m_q/{\overline m}_q$
to ensure the cancellation of the $b\!=\!1/2$ renormalon
contributions. This is so because the coefficient
${\tilde r}_n(\mu)$ contains a polynomial of $n$'th
grade in $\ln[\mu/({\overline m}_q a(\mu))]$ [cf.~Eqs.~(\ref{Eqqexp}),
(\ref{Lp}), (\ref{Eqqs})] which, at large order $n$
and in the large-$\beta_0$ approximation,
sums up approximately to a term 
$\sim (\beta_0/2)^n n! \exp{\ln[\mu/({\overline m}_q a(\mu))]}
= (\beta_0/2)^n n! (\mu/{\overline m}_q) 1/a(\mu)$ \cite{Hoang:1998ng}, 
effectively reducing the power 
of $a(\mu)$ in $E_{q \bar q}(s)/{\overline m}_q$ by one.
Further, the factors $(\beta_0/2)^n$, $n!$ and $\mu$ 
in the approximate sum of the logarithmic terms 
in ${\tilde r}_n(\mu)$ reflect the effect of the
leading ($b\!=\!1/2$) IR renormalon in 
$E_{q \bar q}(s)/{\overline m}_q$.
 
For the Borel-related resummations of $E_{q \bar q}(s)$,
which would account for the leading IR renormalon structure,
we have on the basis of these facts
in principle at least two possible directions to proceed.
The first direction would be to use the Borel transform
of the expansion of $E_{q \bar q}(s)/{\overline m}_q =
\sum_{0}^{\infty} {\tilde r_n}(\mu) a^{n+2}(\mu)$ where the
transformation $a(\mu) \mapsto b$ is performed
literally with respect to all $a(\mu)$-dependence,
including the one appearing in the coefficients
${\tilde r_n}$. This would result in a Borel
transform whose power expansion around the origin
would include terms $b^k \ln^{\ell} b$
with $\ell = 0, 1, 2, \ldots$. 

The second direction
would be to divide the considered quantity by $a(\widetilde \mu)$ 
($\Rightarrow E_{q \bar q}(s)/[{\overline m}_q a(\widetilde \mu)]$), 
where $\widetilde \mu$ is any fixed soft scale,
and then consider the coefficients in the expansion
of this quantity in powers of $a(\mu)$ as independent of
$a(\mu)$, e.g., by expressing them in terms of $a(\widetilde \mu)$.
In the obtained expansion, the coefficients now
contain powers of logarithms $\ln[a(\widetilde \mu)]$ which
are considered as constant (nonvariable) under the
Borel transformation $a(\mu) \mapsto b$.\footnote{
This is in close analogy with the behavior of the static potential
$V_{q \bar q}(r)$ and its dimensionless version
$r V_{q \bar q}(r)$ where 
$r \sim a_{\rm B.} \sim 1/[{\overline m}_q a(\widetilde \mu)]$
(see, for example, 
Refs.~\cite{Hoang:1998nz,Brambilla:1999xf,Beneke:1998rk,Pineda:2001zq,Lee:2003hh,Cvetic:2003wk}).}
It is possile to see that, at large $n$ and in
the large-$\beta_0$ approximation, this is equivalent to the
first approach, because the powers of $a(\mu)$
have been decreased by one, and the coefficients
are now proportional to $(\beta_0/2)^n n! \mu/a(\widetilde \mu)$ where
the factor $1/a(\widetilde \mu)$ is now formally constant
and does not affect the Borel transform (except as an
overall constant factor). The equivalence is assumed
to persist when we go beyond the large-$\beta_0$ approximation,
in the same spirit as the authors of Ref.~\cite{Hoang:1998ng}
assume their conclusions to be valid beyond large-$\beta_0$.

We stress that in both approaches the original expansion of 
$E_{q \bar q}(s)$ in powers of $a(\mu)$ is recovered by applying 
the Borel integration according to the standard formula (\ref{BSint})
term-by-term to the expansion of the Borel transform
around $b\!=\!0$.

In this work, we decide to follow the second direction.
The main reason for this is of practical nature: 
The first approach would generate in the expansion of the
Borel transform around $b\!=\!0$ the terms containing $\ln b, \ln^2 b, \cdots$,
which introduce, at any finite order at least, a cut-singularity 
along the entire negative axis in the $b$-plane. 
We are working at finite orders.
This cut would seriously hamper our re-summations.
For example, the quantity analogous to $R_S(b)$
Eq.~(\ref{RSm}) of the previous Section, but this
time for $E_{q \bar q}(s)/{\overline m}_q$ with the
first approach, has a cut along $b \leq 0$, i.e., 
starting already at the origin, and the resummation at $b\!=\!1/2$
would be difficult. On the other hand, the analogous
quantity $R(b)$ for $E_{q \bar q}(s)/[{\overline m}_q a(\widetilde \mu)]$
in the second approach has no singularities
at $|b| < 1/2$, and for $1/2 \leq |b| < 1$ has
only a cut without infinity along the positive axis. 
Such a quantity can be
much more easily resummed on the basis of its expansion
around $b\!=\!0$. Nonetheless, the first approach presents an 
interesting alternative for which resummation techniques other than
those presented here would have to be developed and/or applied. 

Thus, we will divide the soft binding energy
with the quantity ${\overline \mu}({\widetilde \mu})
= (4/3) {\overline m}_q \alpha_s({\widetilde \mu})$, where
${\widetilde \mu}$ can be any soft scale.
We will fix this scale by the condition
${\widetilde \mu} = (4/3) {\overline m}_q  \alpha_s({\widetilde \mu})$
($\Rightarrow {\widetilde \mu}  = \mu_s$).
Further, in the logarithms $L_p(\mu)$ we
express the pole mass $m_q$ in terms of ${\overline m}_q$
and powers of $a(\mu)$ (cf.~Sec.~\ref{mass}),
and the powers of logarithms $\ln^k[a(\mu)]$ we
re-express in terms of $\ln^k [a(\widetilde \mu)]$.
This then results in the following soft binding energy
quantity $F(s)$ to be resummed
\begin{eqnarray}
F(s) & \equiv &  - \frac{9}{4 \pi^2} 
\frac{E_{q \bar q}(s)}{ {\overline m}_q a(\widetilde \mu) }
= a(\mu) \left[ 1 + a(\mu) f_1 + a^2(\mu) f_2 + a^3(\mu) f_3
+ {\cal O}(a^4) \right] \ ,
\label{Fs}
\end{eqnarray}
where the coefficients $f_j$ depend on $\ln a(\widetilde \mu)$
and on three scales: the renormalization
scale $\mu$ ($\agt m_q \alpha_s$), 
the (fixed) soft scale $\widetilde \mu$,
and ${\overline m}_q$. The coefficient $f_3$ depends,
in addition, on the parameters $\kappa$ (\ref{muf})-(\ref{k30us}),
$\mu_s$, and $a_3$ (\ref{a3nf}).
The coefficients $f_j$ are written explicitly in Appendix \ref{app:sbe}.
The $b=1/2$ renormalon in the quantity $F(s)$ is then of the type
of the renormalon of the pole mass $m_q$ discussed in
the previous Sec.~\ref{mass}.

However, if we divided in Eq.~(\ref{Fs}) by $m_q$ instead of
${\overline m_q}$ and at the same time used in the
resulting $f_j$-coefficients $\ln m_q$, the numerical
resummations of $F(s)$ by methods of Sec.~\ref{Eqq} would
give us values for $E_{q \bar q}(s)$ different usually by
not more than ${\cal O}(10^1 {\rm MeV})$ (we checked this
numerically). We will briefly refer to these approaches later in
this Section as ``pole mass'' approaches. A version of such
pole mass bilocal approach was applied in Ref.~\cite{Lee:2003hh}
for resummation of the unseparated $E_{q \bar q}(s\!+\!us)$. 

The ultrasoft part (\ref{Eqqus}), on the other hand,
has no $b = 1/2$ renormalon. The mass scale used there
should also be renormalon free (${\overline m}_q$).
The renormalization scale $\mu$ there should be adjusted downward
to the typical $us$ scale of the associated process
$\mu \mapsto \mu_{us}$ ($\sim\! m_q \alpha_s^2$)
in order to come closer to a realistic estimate\footnote{
The authors of Ref.~\cite{Recksiegel:2003fm} employed a somewhat
similar idea of using different evaluation methods
for contributions to the spectra of heavy quarkonia from
different regimes (short, intermediate and long-distance).
A similar reasoning was employed, in the context
of high-$T$ QCD, in Ref.~\cite{Cvetic:2002ju}.}
\begin{eqnarray}
E_{q \bar q}(us) & \approx & - \frac{4}{9} {\overline m}_q \pi^2 k_{3,0}(us) a^5(\mu_{us}) \ .
\label{Eqqus2}
\end{eqnarray}

\section{Evaluation of the binding energy}
\label{Eqq}

In this Section we will evaluate the soft part of
the ground state energy for the vector $b {\bar b}$ 
[$\Upsilon (1S)$] and
for the vector and scalar $t {\bar t}$ quarkonium.
In addition, we will estimate the ultrasoft part
of the energy, and will extract the value of the
mass ${\overline m}_b$ from the known mass of $\Upsilon (1S)$.

\subsection{Methods of resummation for the soft energy}
\label{Eqqsmeth}

At first we will apply the same methods as 
those used in Sec.~\ref{mass}.
However, the expansion we will use for the soft energy
quantity $F(s)$ (\ref{Fs}) is higher by one order in
$a(\mu)$ than in quantity $S$ Eq.~(\ref{Sm}) of Sec.~\ref{mass}.
In the ${\rm N}^3{\rm LO}$ coefficient $f_3$ we have
dependence on the approximately known coefficient $a_3$
(\ref{a3nf}), and on the $s$-$us$ factorization scale
parameter $\kappa\!\sim\!1$ Eq.~(\ref{muf}) -- see Appendix
\ref{app:sbe}, Eqs.~(\ref{f3}). It turns out that,
in $f_3$ ($f_3^{(0)}$), the coefficient at $\ln \kappa$ 
is larger than the coefficient at $a_3/(100 \times 4^3)$.
On the other hand, the coefficient at $\ln \kappa$
in the ground state expectation value of the static
potential is about one tenth of the corresponding coefficient
in the (soft) ground binding energy
\begin{subequations}
\label{a3kappa}
\begin{eqnarray}
E_{q \bar q}(s; \ln \kappa\!-\!\text{part}) & \approx &
- 1.93 \times 10^3  \left( {\overline m}_q a^4(\mu) \right) \ln \kappa \ ,
\label{kappa1}
\\
\langle 1 | V_{q \bar q}(r) | 1 \rangle 
(\ln \kappa\!-\!\text{part}) & \approx &
- 1.95 \times 10^2 \left( {\overline m}_q a^4(\mu) \right) \ln \kappa \ ,
\label{kappa2}
\\
E_{q \bar q}(s; a_3\!-\!\text{part}) & = & 
\langle 1 | V_{q \bar q}(r) | 1 \rangle (a_3\!-\!\text{part})
\approx  - 8.77 \times 10^2 
\left( {\overline m}_q a^4(\mu) \right) \frac{a_3}{100 \times 4^3}
\label{a3parts}
\end{eqnarray}
\end{subequations}
Since $a_3/(100 \times 4^3)$ is roughly 
between zero and one [cf.~Eq.~(\ref{a3nf})],
as is also $\ln \kappa$, Eqs.~(\ref{a3kappa}) show that the static potential
is more influenced by the values of $a_3$ than by $\ln \kappa$, while
the situation with the (soft) binding energy is just reversed.
More specifically:
(a) the static potential is more appropriate to obtain
approximate values of $a_3$, as was done e.g.~in Ref.~\cite{Cvetic:2003wk}
and given in Eqs.~(\ref{a3nf}); (b) the soft part of the binding energy
$E_{q \bar q}(s)$ is more appropriate to obtain approximate values
of the $s$-$us$ factorization scale parameter $\kappa$.
We recall that in \cite{Cvetic:2003wk}, the values of $a_3$
(\ref{a3nf}) were obtained by requiring that the known values
of the renormalon residue parameter $N_m$ (\ref{Nmnf})
be reproduced from the Borel transform of the static potential
function $r V_{q \bar q}(r)$. Here we will proceed analogously,
and will obtain approximate values of $\kappa$ (\ref{muf})
by requiring that the residue parameter values (\ref{Nmnf})
be reproduced from the Borel transform of the soft binding
energy quantity $F(s)$ of Eq.~(\ref{Fs}). 

As already mentioned, in contrast to the situation in Sec.~\ref{mass}, the
coefficients $f_j$ of the perturbation series (\ref{Fs})
have some terms proportional to $\ln^k (a(\mu_{*}))$  
($k=1, 2, \ldots$) where $\mu_{*}$ generically denotes fixed chosen 
scales $\widetilde \mu$, or $\mu_s$ -- cf.~Appendix \ref{app:sbe}. 
Here we will argue that these scales should be
between hard and ultrasoft.
These terms are considered constant, independent of $a(\mu)$, 
although they can be formally re-expressed in
terms of $\ln^k a(\mu)$.  
The terms of the type $\ln a$ in the problem at hand are the 
leading terms of logarithms of ratios of various scales
appearing in the problem (cf.~Ref.~\cite{Kniehl:2002br}),
among them $\ln(E_{\rm S}/E_{\rm H})$ and 
$\ln(E_{\rm US}/E_{\rm S})$. The typical hard, soft,
and ultrasoft scales of the problem are, e.g.,
$E_{\rm H} = m_q$, $E_{\rm S} = \langle 1/r \rangle$,
$E_{\rm US} = E_{q {\bar q}}$, i.e., quantities
independent of the renormalization scale ($\mu$).\footnote{
A very similar phenomenon occurs in the perturbation
expansion of the free energy of the high-temperature
quark-gluon plasma, where the hard scale is the
Matsubara frequency $2 \pi T$, and the soft scale
is the Debye screening mass $m_E$ ($\sim\!g_s T$) 
\cite{Cvetic:2002ju,Kajantie:2002wa}.}
The $\mu$-independent ratios of the type $E_{\rm S}/E_{\rm H}$ and
$E_{\rm US}/E_{\rm S}$ have expansions
$E_{\rm X}/E_{\rm Y} = a(\mu) [ 1 + {\cal O}(a) ]$.
The typical resummed value of this quantity can be
written as $a(\mu_{*})$ where $\mu_{*}$ is
the typical scale of the quasiobservable $E_{\rm X}/E_{\rm Y}$.
This suggests that the $\ln a(\mu_{*})$-terms in the coefficients of the
perturbation series should really be somewhere
between hard ($E_{\rm H}\sim\!m_q$) and 
ultrasoft ($E_{\rm US}\sim\!m_q \alpha_s^2$) scales.

Similarly as in Eq.~(\ref{BSrenan}), we have
\begin{eqnarray}
B_{F(s)}(b; \mu) & = & N_m \frac{9}{2 \pi}
\frac{\mu}{ {\overline m}_q a(\widetilde \mu) } 
\frac{1}{ ( 1\!-\!2 b)^{1 + \nu} } \left[ 1 +
\sum_{k=1}^{\infty} {\widetilde c}_k ( 1\!-\!2 b)^k \right]
+ B_{F(s)}^{\rm (an.)}(b; \mu) \ ,
\label{BFrenan}
\end{eqnarray}
where the factor in front of the singular part
was determined by the condition of renormalon cancellation
of the sum $2 m_q + E_{q \bar q}(s)$. We now define in analogy
with Eq.~(\ref{RSm})
\begin{equation}
R_{F(s)}(b; \mu; \mu_f) = (1 - 2 b)^{1 + \nu} B_{F(s)}(b;\mu; \mu_f) \ .
\label{RFs}
\end{equation}
Here we denoted, for clarity, explicitly the dependence 
on the factorization scale $\mu_f$.
Expressions (\ref{BFrenan}) and (\ref{RFs}) imply
\begin{equation}
N_m = \frac{2 \pi}{9} \frac{ {\overline m}_q a(\widetilde \mu) }{\mu}
R_{F(s)}(b;\mu;\mu_f) {\big |}_{b=1/2} \ .
\label{NmEqq}
\end{equation}
The expansion of $R_{F(s)}$ is exactly known up to $\sim\!b^2$,
and approximately up to $\sim\!b^3$ (${\rm N}^3{\rm LO}$ TPS), 
where the latter coefficient
is dependent on $\kappa$ (and, more weakly, on $a_3$). All coefficients
are dependent also on the renormalization scale $\mu$ 
($\agt m_q \alpha_s$). It turns out that the expansion
(\ref{NmEqq}) is significantly less convergent than
the series (\ref{RSm}) (at $b=1/2$). However, it is not
clearly divergent, unless we take unreasonable values of $\kappa$ or
$\mu$. Theoretically, $R_{F(s)}(b)$ should be a function
with only a weak singularity (cut) at $b=1/2$, and the
nearest pole at $b=3/2$ (i.e., the next renormalon pole of
$V_{q \bar q}(r)$ \cite{Aglietti:1995tg}). 
Thus, resummations such as Pad\'e approximations (PA's)
are expected to work better on $R_{F(s)}(b)$ than on
$B_{F(s)}(b)$. The Pad\'e approximation with the simplest
pole structure for the ${\rm N}^3{\rm LO}$ TPS is $[2/1]$, i.e.,
ratio of a quadratic with a linear polynomial in $b$.
It turns out that $R_{F(s)}[2/1](b)$ has physically acceptable
pole structure $|b_{\rm pole}| \geq 1$ for most of the
values of $\mu \agt m_q \alpha_s$ and $\kappa \sim 1$.
Using this Pad\'e to evaluate expression (\ref{NmEqq})
gives us predictions for the residue parameter $N_m$
reasonably stable under the variation of $\mu$. On the other
hand, the predicted value of $N_m$ depends significantly
on the $s$-$us$ factorization scale parameter $\kappa$
(\ref{muf}). 

\begin{figure}[htb]
\begin{minipage}[b]{.49\linewidth}
 \centering\epsfig{file=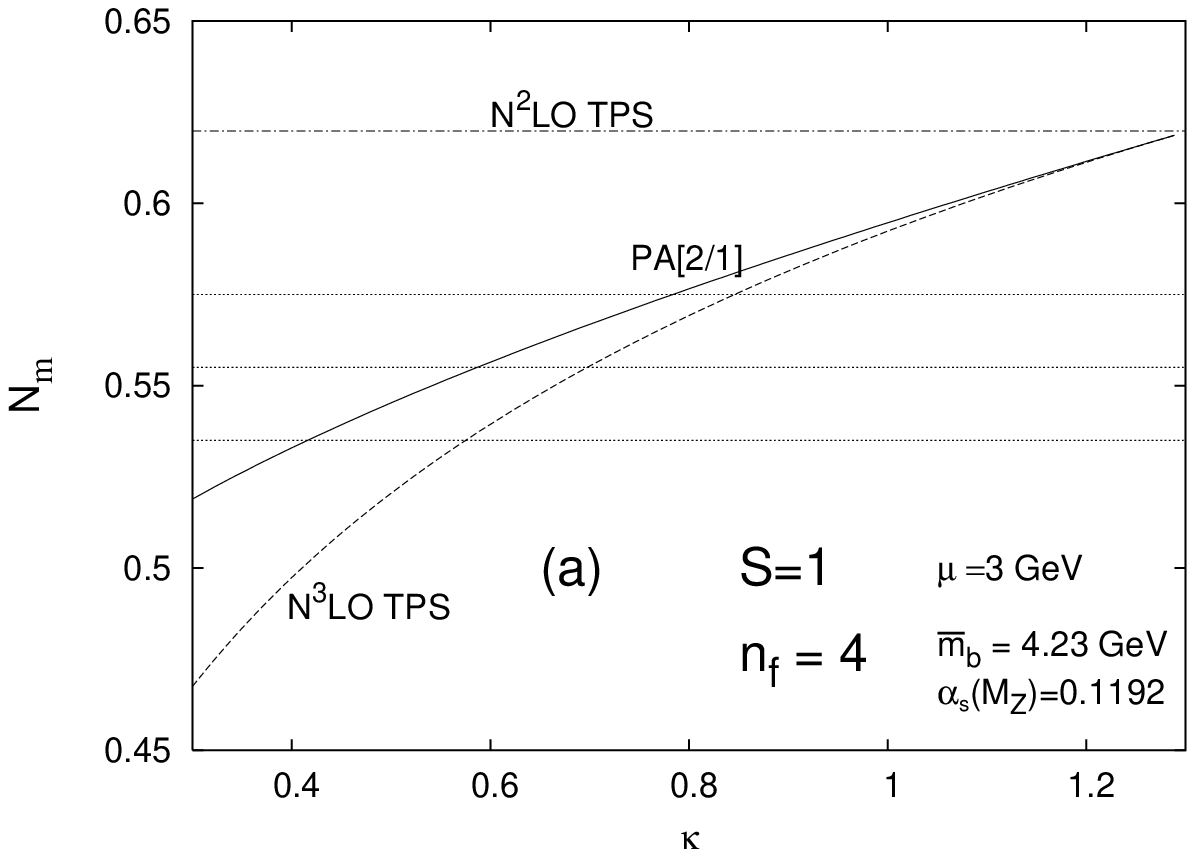,width=\linewidth}
\end{minipage}
\begin{minipage}[b]{.49\linewidth}
 \centering\epsfig{file=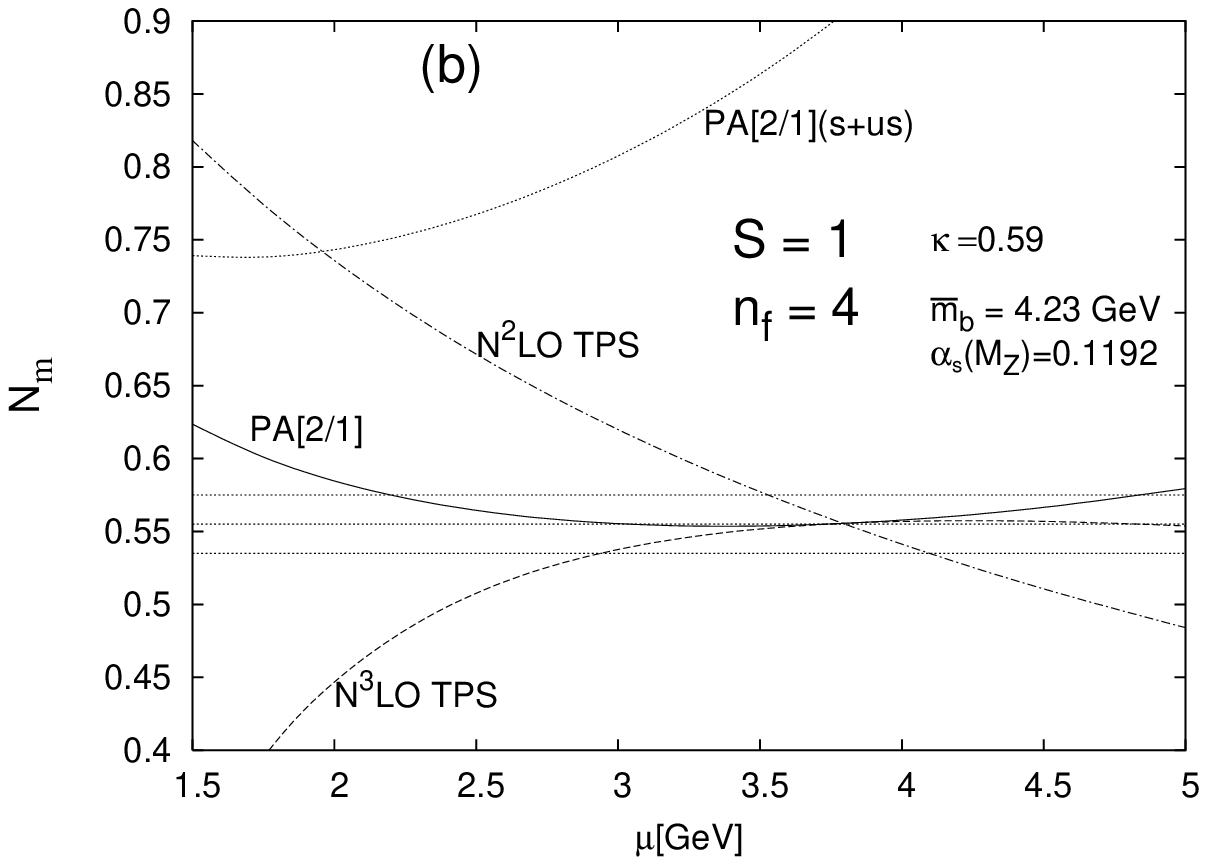,width=\linewidth}
\end{minipage}
\vspace{0.2cm}
\caption{The residue parameter value $N_m$ as calculated from the
soft part of the binding energy of the bottonium according to 
Eq.~(\ref{NmEqq}), (a) as a function of the $s$-$us$ factorization
scale parameter $\kappa$ (\ref{muf}), at $\mu=3$ GeV; 
(b) as a function of the renormalization scale $\mu$, at
$\kappa = 0.59$. Further explanations given in the text.
In Fig.~(a), the known values (\ref{Nmnf4}) of $N_m$ are denoted as dotted
horizontal lines.}
\label{Nmbb.fig}
\end{figure}

In Fig.~\ref{Nmbb.fig}(a) we show the
dependence of $N_m$ on $\kappa$, at a typical (``central'')
$\mu$ value $\mu\!=\!3$ GeV, for the $b \bar b$ system.
The known central value (\ref{Nmnf4}) of $N_m$ is
obtained by the $R_{F(s)}[2/1](b\!=\!1/2)$ expression at
$\kappa \approx 0.59$. In Fig.~\ref{Nmbb.fig}(b) we
present, for $\kappa = 0.59$, the dependence of
calculated $N_m$ on the renormalization scale $\mu$.
There, we include also the ($[2/1]$-resummed) curve for
the case when no separation of the $s$ and $us$ parts
of the energy is performed. In that case, the
obtained values of $N_m$ are unacceptable.
If the ``pole mass'' version is applied
[mentioned in the second paragraph after Eq.~(\ref{Fs})],
with no separation of the $s$ and $us$ parts,
the obtained values of the ($[2/1]$-resummed) curve 
remain above $0.70$ as well, thus unacceptable.
The other values of the input parameters are chosen
to have the $b \bar b$ ``central'' values: 
$a_3/4^3 = 86$ (\ref{a3nf4}); ${\overline m}_b = 4.23$ GeV;
${\widetilde \mu} = 1.825$ GeV ($\approx \mu_s$)
and $\alpha_s(\widetilde \mu; n_f\!=\!4) = 0.3263 (\approx 
\alpha_s(\mu_s; n_f\!=\!4) = 0.326$) 
[from: $\alpha_s(m_{\tau}; n_f\!=\!3) = 0.3254$, i.e.,
$\alpha_s(M_Z) = 0.1192$ \cite{Cvetic:2001ws}].
For the RGE running, we always use four-loop ${\overline {\rm MS}}$
$\beta$-function (TPS) and three-loop quark threshold matching
relations \cite{Chetyrkin:1997sg}, with $\mu_{\rm thresh.} = 
2 {\overline m}_c$, $ 2 {\overline m}_b$.
\begin{figure}[htb]
\begin{minipage}[b]{.49\linewidth}
 \centering\epsfig{file=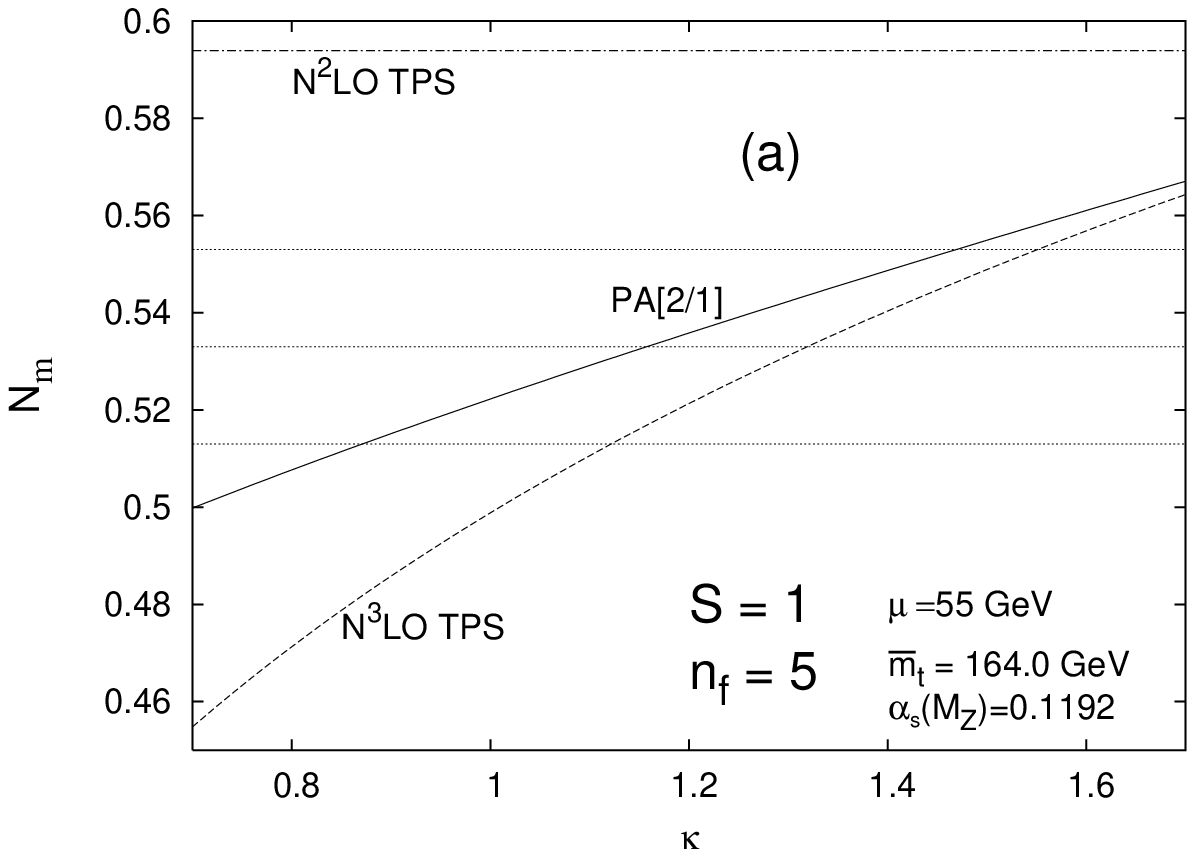,width=\linewidth}
\end{minipage}
\begin{minipage}[b]{.49\linewidth}
 \centering\epsfig{file=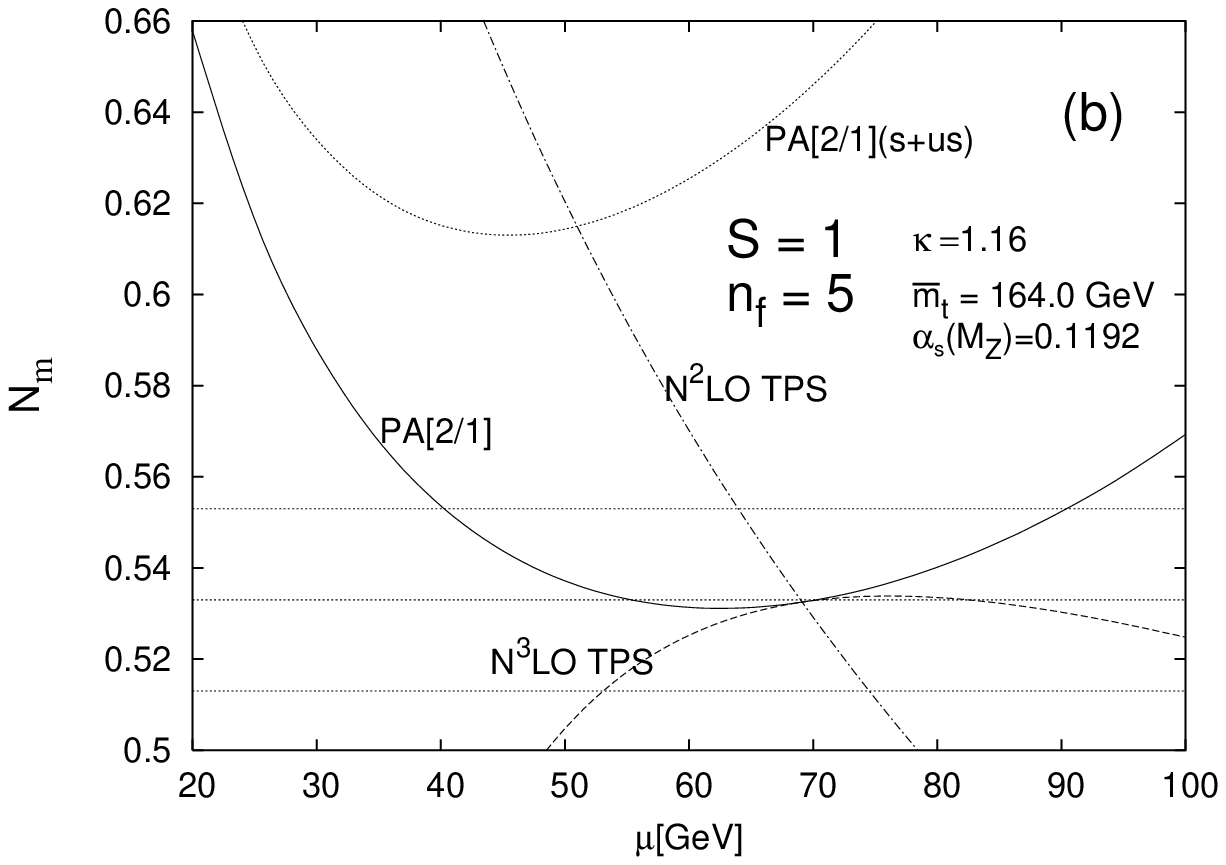,width=\linewidth}
\end{minipage}
\vspace{0.2cm}
\caption{Same as in Fig.~\ref{Nmbb.fig}, but for the
($S=1$) toponium. In Fig.~(a), $\mu = 55$ GeV and
the known values (\ref{Nmnf5}) of $N_m$ are given as dotted horizontal lines.}
\label{NmttS1.fig}
\end{figure}
\begin{figure}[htb]
\begin{minipage}[b]{.49\linewidth}
 \centering\epsfig{file=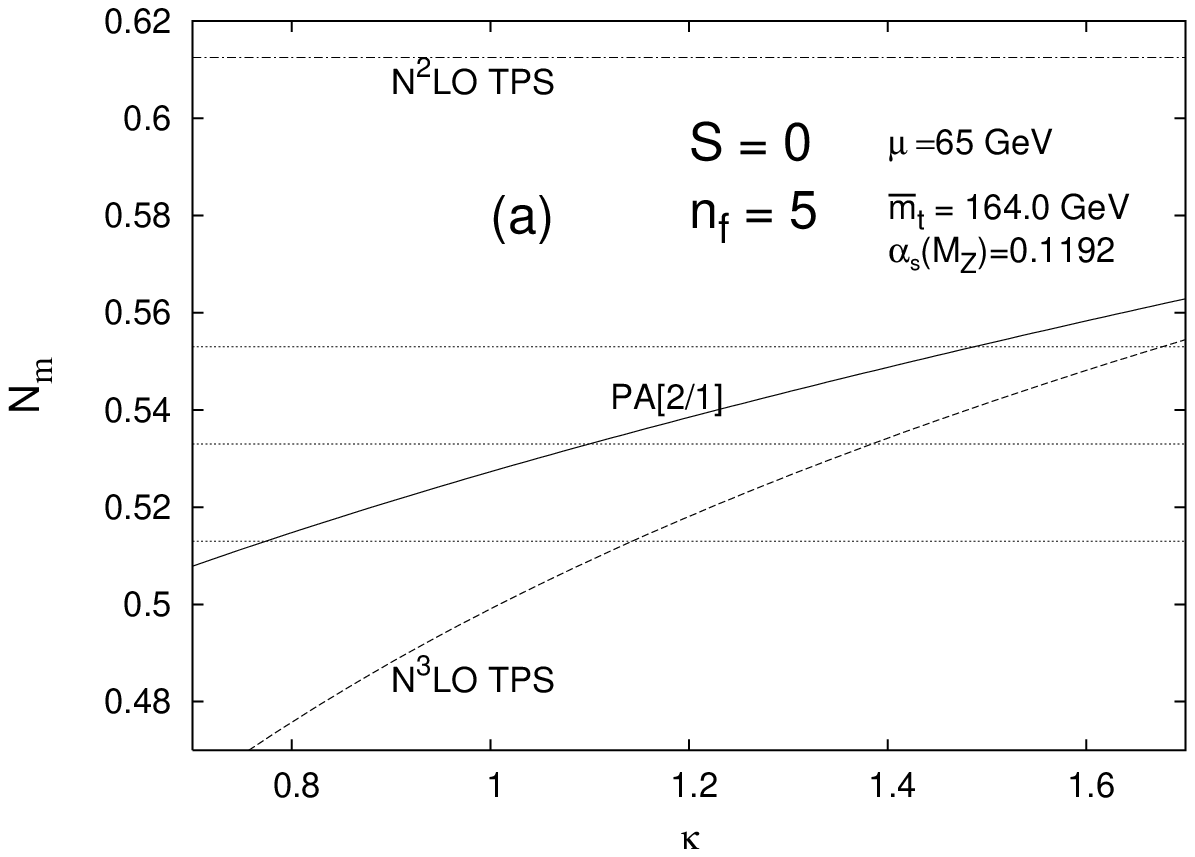,width=\linewidth}
\end{minipage}
\begin{minipage}[b]{.49\linewidth}
 \centering\epsfig{file=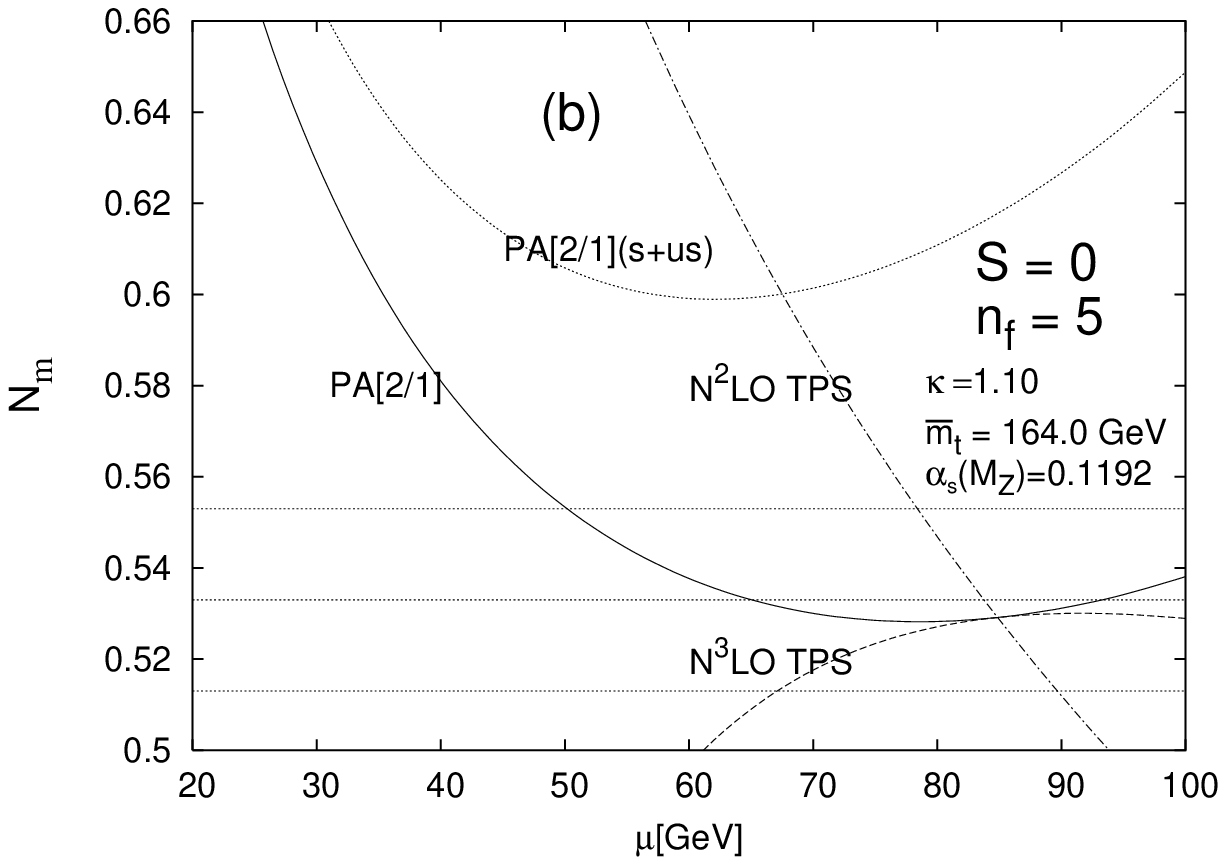,width=\linewidth}
\end{minipage}
\vspace{0.2cm}
\caption{Same as in Fig.~\ref{NmttS1.fig}, but for the scalar ($S=0$)
toponium. In Fig.~(a), $\mu = 65$ GeV is taken.}
\label{NmttS0.fig}
\end{figure}

In Figs.~\ref{NmttS1.fig} and \ref{NmttS0.fig} we present analogous
results for the $t \bar t$ vector ($S=1$) and scalar ($S=0$)
bound state. The typical (``central'')
values of the renormalization scale were chosen to be
$\mu = 55$ GeV and $65$ GeV, respectively. The
$s$-$us$ factorization parameter $\kappa$ values obtained
were $\kappa = 1.16$ ($S=1$) and $\kappa=1.10$ ($S=0$),
so that $R_{F(s)}[2/1](b=1/2)$ would reproduce the
known central value (\ref{Nmnf5}) of the residue parameter $N_m$.
The other input parameters have the $t \bar t$
``central'' values:
$a_3/4^3 = 62.5$ (\ref{a3nf5}); 
${\overline m}_t = 164.0$ GeV;
${\widetilde \mu} = 31.$ GeV ($\approx \mu_s$)
and $\alpha_s(\widetilde \mu; n_f\!=\!5) = 0.1430$ [$\approx 
\alpha_s(\mu_s; n_f\!=\!5) = 0.14$]. 
The values of $N_m$ extracted when no separation
of $s$ and $us$ is performed, are unacceptably high
$N_m \geq 0.6$ (also in the ``pole mass'' version: $N_m \geq 0.6$). 

Variation $N_m = 0.555 \pm 0.020$ [$b \bar b$, Eq.~(\ref{Nmnf4})]
implies $\kappa = 0.59 \pm 0.19$; variation
$N_m = 0.533 \pm 0.020$ [$t \bar t$, Eq.~(\ref{Nmnf5})]
implies $\kappa = 1.16^{\ + \ 0.31}_{\ - \ 0.29}$ ($S\!=\!1$) and
$\kappa = 1.10^{\ + \ 0.39}_{\ - \ 0.33}$ ($S\!=\!0$).
If, on the other hand, $a_3$ parameter is varied,
according to Eqs.~(\ref{a3nf}), then for $b \bar b$
$\kappa = 0.59 \mp 0.06$, 
and for $t \bar t$
$\kappa = 1.16^{\ -\ 0.10}_{\ +\ 0.11}$ ($S\!=\!1$) and
$\kappa = 1.10^{\ - \ 0.09}_{\ + \ 0.11}$ ($S\!=\!0$).
Thus, the value of $s$-$us$ factorization scale parameter $\kappa$
is influenced largely by the allowed values (\ref{Nmnf})
of the renormalon residue parameter, and significantly less
by the allowed values $a_3$ (\ref{a3nf}) 
of the ${\rm N}^3{\rm LO}$ coefficient of the static $q \bar q$
potential. Therefore, we will consider the variations
of $N_m$ (\ref{Nmnf}) and of $\kappa$ to be 
related by a one-to-one relation, while the variations
of $a_3$ (\ref{a3nf}) will be considered as independent.

In this way, we have the following values for the $s$-$us$ 
factorization scale parameter:
$\kappa$ (\ref{muf})
\begin{subequations}
\label{kapnf}
\begin{eqnarray}
N_m = 0.555 \pm 0.020 \ \Rightarrow \
\kappa & = & 0.59 \pm 0.19 
\quad (n_f = 4, \ S = 1) \ ,
\label{kapnf4}
\\
N_m = 0.533 \pm 0.020 \ \Rightarrow \
\kappa & = & 1.16^{\ + \ 0.31}_{\ - \ 0.29} 
\quad (n_f = 5, \ S = 1) \ ,
\label{kapnf5S1}
\\
 \ \Rightarrow \
\kappa & = & 1.10^{\ + \ 0.39}_{\ - \ 0.33} 
 \quad (n_f = 5, \ S = 0) \ ,
\label{kapnf5S0}
\end{eqnarray}
\end{subequations}
and thus we obtain the ${\rm N}^3{\rm LO}$ TPS (\ref{Fs})
for the soft part of the ground binding energy.

We wish to add a comment on $\kappa$-dependence of $N_m$ in 
Figs.~\ref{Nmbb.fig}-\ref{NmttS0.fig}. Theoretically,
the parameter $N_m$ should be independent of the
$s$-$us$ factorization scale $\mu_f$ and thus
independent of the related parameter $\kappa$ (\ref{muf}).
However, among the $\mu_f$-dependent terms in
$R_{F(s)}(b;\mu;\mu_f)$ of Eq.~(\ref{RFs}), only the leading term 
is available. Due to this restrictive practical situation, 
the value of the residue parameter $N_m$ obtained by Eq.~(\ref{NmEqq})
automatically possesses significant $\mu_f$-dependence
(or: $\kappa$-dependence), and the value of $\mu_f$ (or: $\kappa$)
is fixed by requiring that this leading order
expression in $\mu_f$ reproduce the known value of $N_m$.
The value of $\mu_f$ obtained in this way must be physically acceptable
($\Leftrightarrow \kappa\!\sim\!1$) if the procedure is consistent.
This is analogous to the situation when a QCD observable
$S(Q)$ is known to the leading order $\sim\!a(\mu)$ only. Equating
such leading order expression $S_{[1]}(Q;\mu)$ with the
known value of $S(Q)$, a specific value of the
renormalization scale $\mu = \mu_{\rm ECH}$
is obtained such that $S_{[1]}(Q;\mu_{\rm ECH}) = S(Q)$.
This is the main idea of the effective charge (ECH) method \cite{ECH}. 
If the procedure is consistent, the obtained renormalization scale
$\mu$ value $\mu_{\rm ECH}$ should be of the order of the
physical scale $Q$ of the process associated with the
observable: $\mu_{\rm ECH}/Q \sim 1$. The analogy
with our case consists in the correspondence
$\mu_f \leftrightarrow \mu$, 
$E_{\rm US} (\sim\!E_{q \bar q}) \leftrightarrow Q$,
and $\mu_f(\rm obtained) \leftrightarrow \mu_{\rm ECH}$.

Now that the value of $\kappa$ has been obtained,
and consequently the ${\rm N}^3{\rm LO}$ TPS (\ref{Fs}),
we can perform the resummation of the soft part of the
ground binding energy.
The full bilocal method \cite{Lee:2002sn,Lee:2003hh}
can be performed as in Sec.~\ref{mass},
Eqs.~(\ref{BSbiloc}) and (\ref{BSint}). However,
now we have one term more in the TPS. Therefore
\begin{eqnarray}
B_{F(s)}^{(\rm biloc.)}(b;\mu) & = & 
N_m \frac{9}{2 \pi}
\frac{\mu}{ {\overline m}_q a(\widetilde \mu) } 
\frac{1}{ ( 1\!-\!2 b)^{1 + \nu} } \sum_{k=0}^3 {\widetilde c}_k (1 - 2 b)^k
+ \sum_{k=0}^3 \ \frac{h_k}{k! \ \beta_0^k} \ b^k \ ,
\label{Fsbiloc}
\end{eqnarray}
where the coefficients ${\widetilde c}_k$ are given by Eqs.~(\ref{tc})
and (\ref{tr})  (${\widetilde c}_0 = 1$), and the coefficients
$h_k$ in the expansion of the analytic part are now known
up to oder $k=3$
\begin{equation}
h_k = f_k - N_m \ \frac{9}{2 \pi} \ 
\frac{\mu}{ {\overline m}_q a({\widetilde \mu}) } \
(2 \beta_0)^k \sum_{n=0}^3 \ 
{\widetilde c}_n 
\frac{ \Gamma ( \nu + k + 1 - n) }{ \Gamma(\nu + 1 - n) } 
\quad (k=0,1,2,3) \ .
\label{hss}
\end{equation}
Here, by convention, $f_0 = 1 = {\widetilde c}_0$. 
Then the resummed quantity is obtained
by taking the PV of the Borel integration of $B_{F(s)}(b)$ of 
Eq.~(\ref{Fsbiloc}), as in Sec.~\ref{mass} for $B_S(b)$
[Eq.~(\ref{BSint}), integration along a ray]. The result
would have some spurious $\mu$-dependence. However, for the
typical $\mu$-scales $m_q \agt \mu \agt m_q \alpha_s$, the
analytic part $B_{F(s)}^{\rm (an.)}(b)$ of the Borel transformation 
in Eq.~(\ref{Fsbiloc}) turns out to have a problematic behavior 
in the following sense. When it is Pad\'e-resummed
as $B_{F(s)}^{\rm (an.)}[2/1](b)$, the obtained pole is almost always
(for most $\mu$'s) unacceptably small in size: 
$|b_{\rm pole}| \leq 1/2$. Theoretically, $B_{F(s)}^{\rm (an.)}(b)$
should have the nearest pole at $b=3/2$ \cite{Aglietti:1995tg}. Thus, 
$B_{F(s)}^{\rm (an.)}$ appears to be too singular in the
above bilocal approach, and the TPS and Pad\'e evaluations
of it would result in widely differing resummed values for
the energy $E_{q \bar q}(s)$. The reason for this problem
appears to lie in the specific truncated form of the
singular part taken in the bilocal method (\ref{Fsbiloc}).
While the latter part describes well the behavior of the
transform near $b=1/2$, it influences apparently strongly
the coefficients $h_k$ and thus the analytic part,
so that no reliable resummation of that part (apart from
TPS) can be done. In this context, we note that
the series of terms $\sum_k {\widetilde c}_k (1 - 2 b)^k$
has no indication of convergence at $b=0$, as seen from the
values of ${\widetilde c}_j$ in Table \ref{table1} of
Sec.~\ref{mass}. This problem can be alleviated by
introducing in the renormalon part a ``form'' factor
which suppresses that part away from $b \approx 1/2$,
but keeps it unchanged at $b \approx 1/2$. If we choose
for this factor a Gaussian type of function, we are led to
the following set of ``$\sigma$-regularized''
bilocal expressions for the Borel transform
\begin{eqnarray}
B_{F(s)}^{(\sigma)}(b;\mu) & = & 
N_m \ \frac{9}{2 \pi} \
\frac{\mu}{ {\overline m}_q a(\widetilde \mu) } \
\frac{1}{ ( 1\!-\!2 b)^{1 + \nu} } \left[ 
1 + {\widetilde c}_1 (1\!-\!2b) + 
\left( {\widetilde c}_2 + \frac{1}{8 \sigma^2} \right)
(1\!-\!2b)^2 + 
\left( {\widetilde c}_3 + \frac{{\widetilde c}_1}{8 \sigma^2} \right)
(1\!-\!2b)^3
\right] 
\nonumber\\
&&
\times
\exp \left[ - \frac{1}{8 \sigma^2} (1\!-\!2 b)^2 \right] 
+ \sum_{k=0}^3 \ \frac{1}{k! \ \beta_0^k} \ h^{(\sigma)}_k  b^k \ .
\label{Fssig}
\end{eqnarray}
The corrective terms $1/(8 \sigma^2)$ and 
$ {\widetilde c}_1/(8 \sigma^2)$ in the 
coefficients of the renormalon part
of Eq.~(\ref{Fssig}) appear to ensure the correct known behavior
of the renormalon part up to order $\sim\!(1 - 2b)^{-\nu + 2}$.
The coefficients $h_k^{(\sigma)}$ in Eq.~(\ref{Fssig})
differ from $h_k$'s of the bilocal case (\ref{hss}),
and are determined by the requirement that the power
expansion of expression (\ref{Fssig}) reproduce the
known ${\rm N}^3{\rm LO}$ TPS of the Borel transform
of $F_s$ (\ref{Fs}).
If $\sigma$ parameter increases (i.e., $\sigma \agt 1$), 
formula (\ref{Fssig}) is expected to gradually reduce to
the bilocal formula (\ref{Fsbiloc}). If $\sigma \to 0$,
then the expansion of the Gaussian form function in (\ref{Fssig})
would imply very large coefficients ($\agt \sigma^{-4}$)
at the renormalon terms
$\sim\!(1\!-\!2b)^{-1 -\nu + k}$ ($k=4, 5, \ldots$). This
is not expected to reflect the reality, because the results
in Table \ref{table1} suggest that $|{\widetilde c}_k| \alt 1$
for $k = 4, 5, \ldots$. Therefore, we expect that the optimal
choice of $\sigma$ would be somewhere between zero and one.
Numerical analysis confirms this expectation.
Namely, when $\sigma$ decreases from $\sigma\!=\!\infty$ to about
$\sigma \approx 0.3$-$0.4$, the
value of the pole of the $[2/1]$ Pad\'e-resummed analytic part
$B_{F(s)}^{({\rm an.} \sigma)}(b)$ of Eq.~(\ref{Fssig})
gradually turns acceptable ($|b_{\rm pole}| > 1$) and
rather stable when the renormalization
scale $\mu$ varies in the interval $[m_q \alpha_s, m_q]$
(except close to $\mu \approx m_q \alpha_s$). Further,
the Borel resummation with the TPS-evaluated and 
with Pad\'e-evaluated analytic parts give
for such $\sigma$'s similar values of $E_{q \bar q}(s)$, 
indicating that the analytic part now manifests more clearly
its non-singular behavior. When the value of $\sigma$
falls below $0.3$, the analytic part starts showing
erratic behavior again and the Borel resummation
gives significantly differing results with the TPS-
and the Pad\'e-evaluated analytic parts. Further, the
$\sigma$--dependence of the obtained soft energy becomes very strong
for $\sigma < 0.3$.
On these grounds,
the obtained optimal $\sigma$ turn out to be
\begin{subequations}
\label{sig}
\begin{eqnarray}
\sigma &=& 0.36 \pm 0.03 \quad (n_f\!=\!4, \ S=1) \ ,
\label{signf4}
\\
\sigma &=& 0.33 \pm 0.03 \quad (n_f\!=\!5, \ S=0,1) \ .
\label{signf5}
\end{eqnarray}
\end{subequations}
\begin{figure}[htb]
\begin{minipage}[b]{.49\linewidth}
 \centering\epsfig{file=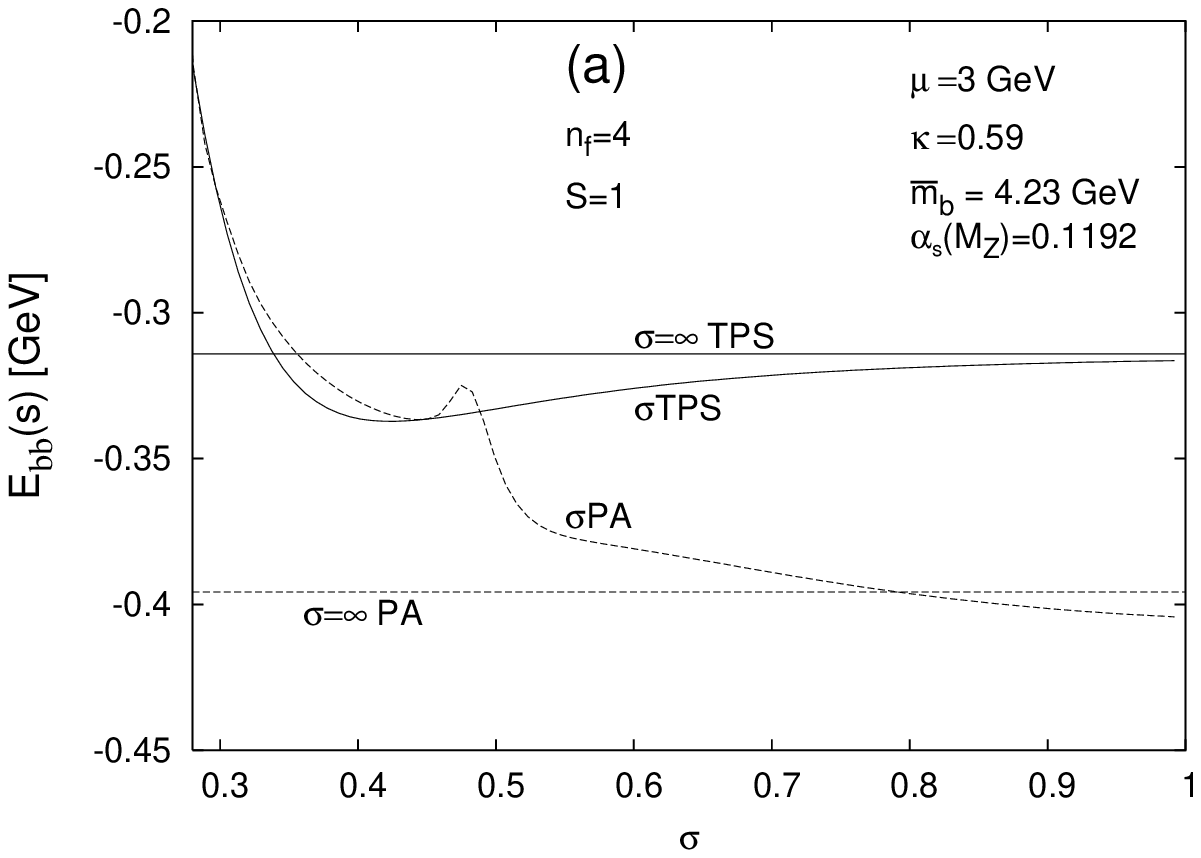,width=\linewidth}
\end{minipage}
\begin{minipage}[b]{.49\linewidth}
 \centering\epsfig{file=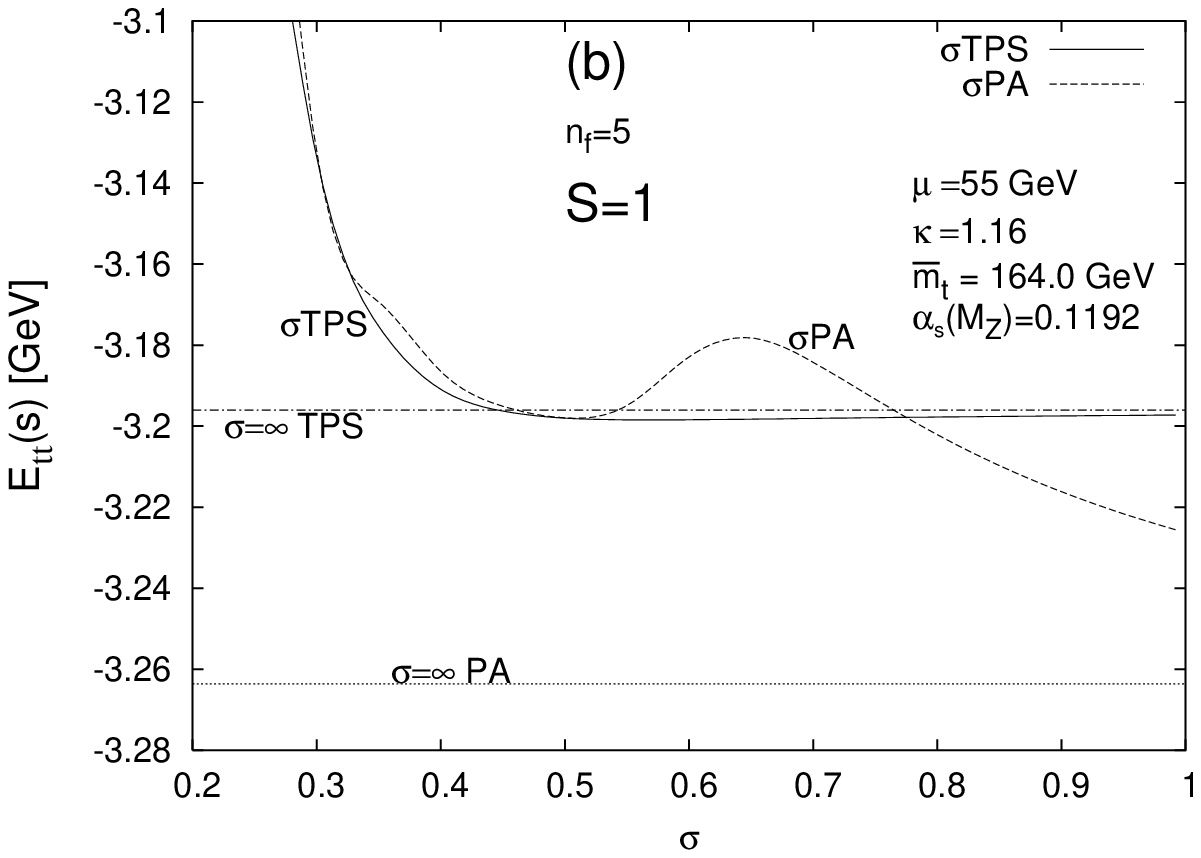,width=\linewidth}
\end{minipage}
\vspace{0.2cm}
\caption{(a) Soft part of the ground state binding energy of
$b {\bar b}$, evaluated with the (PV) Borel-resummed
expression (\ref{Fssig}), as a function of the
method parameter $\sigma$. 
(b) Same as in (a), but for the toponium
$S=1$ system. Details are given in the text.}
\label{Eqqsig.fig}
\end{figure}
In Fig.~\ref{Eqqsig.fig}(a) we present the (PV) Borel-resummed
soft part of ground state energy for the bottonium ($S=1$),
as a function of the $\sigma$ parameter of method (\ref{Fssig}).
The results are given when the
analytic part of Borel transform (\ref{Fssig}) is either
evaluated as ${\rm N}^3{\rm LO}$ TPS or as $[2/1]$ Pad\'e (PA).
In addition, the two corresponding results (TPS, and PA)
are given as horizontal lines when the bilocal method (\ref{Fsbiloc}) 
is applied ($\sigma=\infty$). The values of the other input
parameters have the same 
``central'' values as in Figs.~\ref{Nmbb.fig}, and
$N_m = 0.555$ and $c_4 = 40.$ in accordance with Eqs.~(\ref{Nmnf4})
and (\ref{c4nf4}). 
In Fig.~\ref{Eqqsig.fig}(b) we present
analogous results for the toponium vector ($S=1$) 
soft binding energy. The values of the input
parameters are the same as in Figs.~\ref{NmttS1.fig}
and \ref{NmttS0.fig}, and in addition $N_m = 0.533$ and
$c_4 = 70$ in accordance with Eqs.~(\ref{Nmnf5}) and (\ref{c4nf5}).
The corresponding curves for the toponium scalar ($S\!=\!0$) case 
are very similar to those of the $S\!=\!1$ case.

In addition to the methods (\ref{Fsbiloc}) and (\ref{Fssig})
employed up to now, which are mutually related, we want to
emply as a cross check of our numerical results also a method
unrelated to the (full) bilocal method. This will be the
$R$-method \cite{Caprini:1998wg,Cvetic:2001sn}, where
we resum the function $R_{F(s)}(b;\mu)$ (\ref{RFs})
and then employ the (PV) Borel resummation as written in
Eq.~(\ref{BRSint}) (with $R_{F(s)}$ instead of $R_S$ there).
Since we know the ${\rm N}^3 {\rm LO}$ TPS of $R_{F(s)}(b)$,
we can evaluate this function as TPS, or as Pad\'e
$[2/1]$ (the Pad\'e $[1/2]$ is disfavored due to a more
complicated and unstable pole structure).

\begin{figure}[htb]
\begin{minipage}[b]{.49\linewidth}
 \centering\epsfig{file=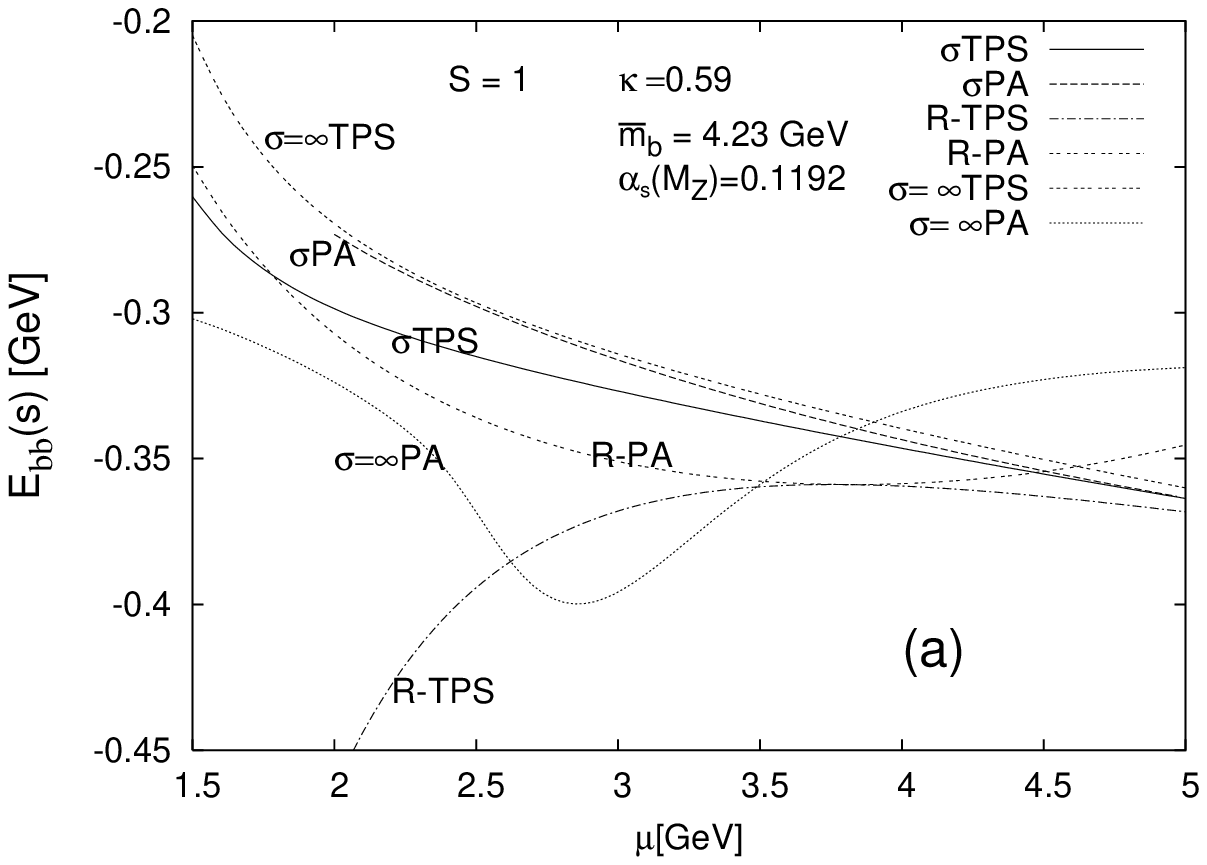,width=\linewidth}
\end{minipage}
\begin{minipage}[b]{.49\linewidth}
 \centering\epsfig{file=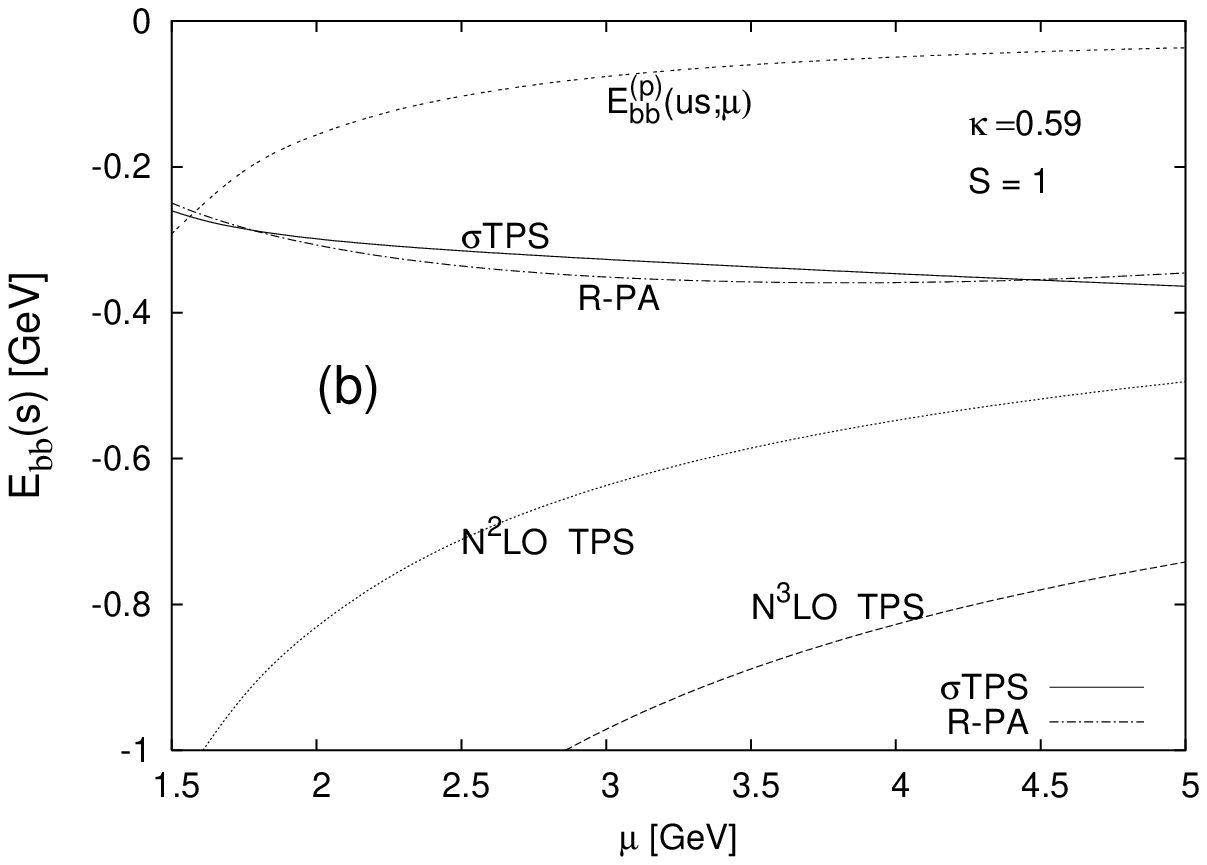,width=\linewidth}
\end{minipage}
\vspace{0.2cm}
\caption{(a) Soft part $E_{b \bar b}(s)$
of the ground state binding energy of
$b {\bar b}$, evaluated with four different methods
involving (PV) Borel resummation, as functions of the
renormalization scale $\mu$. Details are given in the text.
In Fig.~(b) the simple TPS results for $E_{b \bar b}(s)$ 
are included [Eq.~(\ref{FsTPS})],
as well as the ``perturbative'' ultrasoft part
$E_{b \bar b}^{\rm (p)}(us; \mu)$ [Eq.~(\ref{Fusp})].}
\label{Ebbmu.fig}
\end{figure}
The results for the soft binding energy $E_{b \bar b}(s)$
of the ground state of
bottonium, as functions of the renormalization scale $\mu$,
are presented in Fig.~\ref{Ebbmu.fig}(a). The values of input 
parameters are taken as in Figs.~\ref{Nmbb.fig} and \ref{Eqqsig.fig}(a),
and for the ``${\sigma}$-regularized'' method
we take $\sigma=0.36$ according to Eq.~(\ref{signf4})
(note that the $R$-method does not need $N_m$, $c_4$,
and $\sigma$ as input).
For each of the three methods, we present two curves:
when the analytic part is evaluated as TPS, or as Pad\'e
$[2/1]$ (PA), where the role of the analytic part in the $R$-method
is taken over by the function $R_{F(s)}(b)$ itself.
We observe from the Figure that the bilocal method (\ref{Fsbiloc})
($\sigma\!=\!\infty$) gives the TPS and PA results which
significantly differ from each other. On the other hand,
the ``$\sigma$-regularized'' method (\ref{Fssig}) ($\sigma=0.36$)
gives the TPS and PA results closer to each other.
The methods $\sigma$-TPS, $\sigma$-PA, and $R$-PA give
similar results in the entire presented $\mu$-interval.
$R$-TPS appears to fail at low $\mu$ ($\approx m_b \alpha_s
\approx 1$-$2$ GeV).
In Fig.~\ref{Ebbmu.fig}(b) we include, for comparison,
the simple TPS evaluation of $E_{b \bar b}(s)$, according to
formula [cf.~Eq.~(\ref{Fs})]
\begin{equation}
F(s)^{\rm (TPS)}  \equiv -\frac{9}{4 \pi} 
\frac{1}{{\overline m}_b \alpha_s(\widetilde \mu)} E_{q \bar q}(s)
= a(\mu) \left[ 1 + a(\mu) f_1 + a^2(\mu) f_2 + a^3(\mu) f_3 \right]
\ ,
\label{FsTPS}
\end{equation}
where for ${\rm N}^2{\rm LO}$ TPS case we take $f_3=0$.
In Fig.~\ref{Ebbmu.fig}(b) the same input parameters are used 
as in Fig.~\ref{Ebbmu.fig}(a).
We see that the perturbation series shows strongly divergent behavior
already at ${\rm N}^3{\rm LO}$.
In this Figure, we also included the ``perturbative''
ultrasoft part $E_{b \bar b}^{\rm (p)}(us;\mu)$ calculated
according to [see Eqs.~(\ref{k30us}) and (\ref{Eqqus})]. 
\begin{equation}
F^{\rm (p)}(us) \equiv -\frac{9}{4 \pi} 
\frac{1}{{\overline m}_b \alpha_s(\widetilde \mu)} 
E_{q \bar q}^{\rm (p)}(us;\mu) = k_{3,0} a^4(\mu) \ .
\label{Fusp}
\end{equation}
This quantity is highly $\mu$--dependent. We return
to the discussion of the $us$ energy part in the
next Subsection.

\begin{figure}[htb]
\begin{minipage}[b]{.49\linewidth}
 \centering\epsfig{file=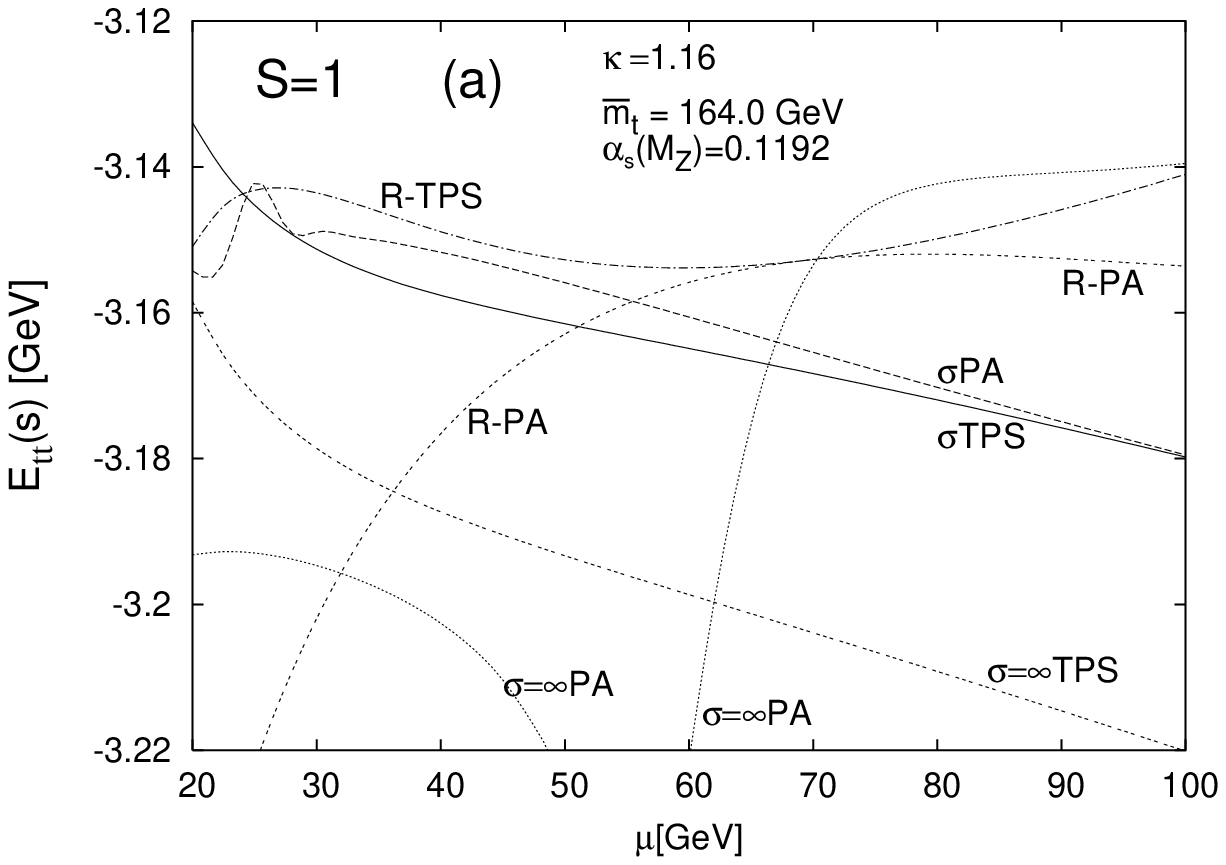,width=\linewidth}
\end{minipage}
\begin{minipage}[b]{.49\linewidth}
 \centering\epsfig{file=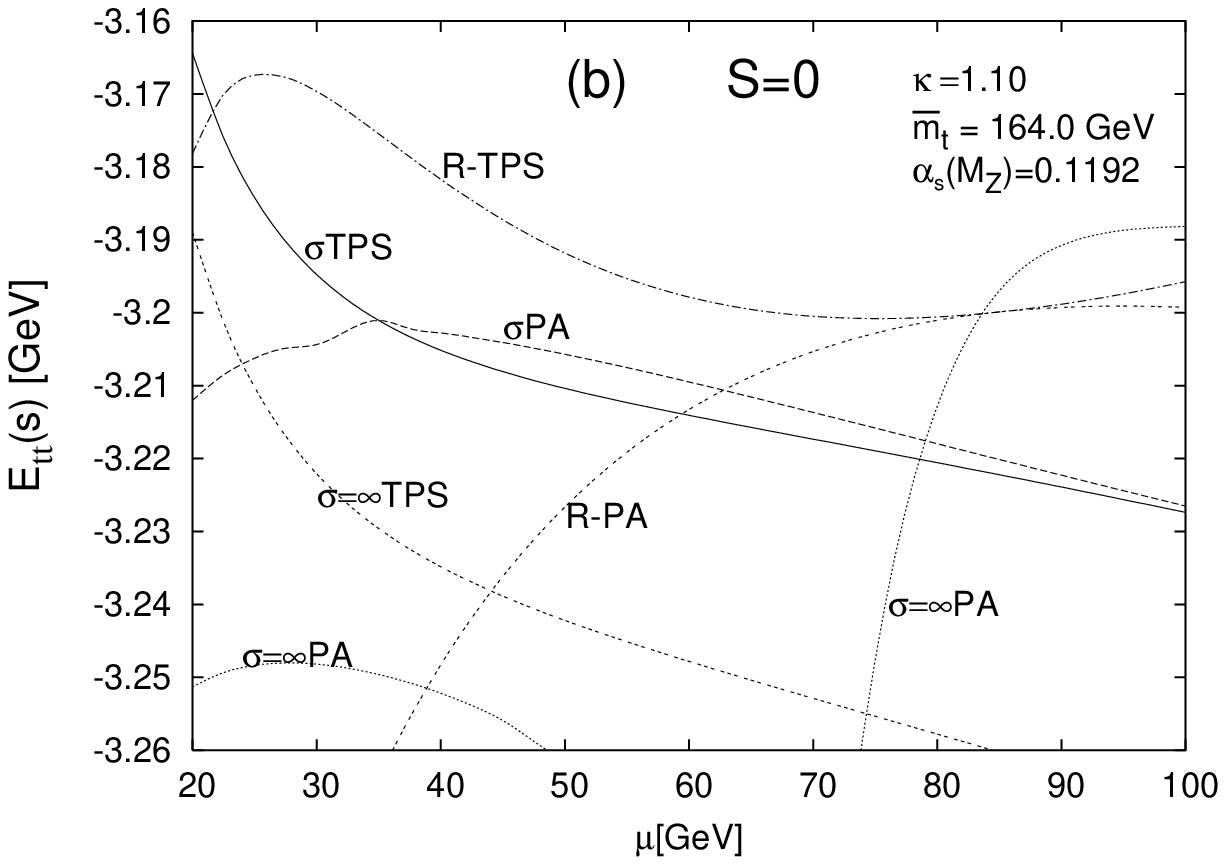,width=\linewidth}
\end{minipage}
\vspace{0.2cm}
\caption{Same as in Fig.~\ref{Ebbmu.fig}(a), but for the toponium
system -- (a) vector ($S=1$), (b) scalar ($S=0$). Details are
given in the text.}
\label{Ettmu.fig}
\end{figure}
In Figs.~\ref{Ettmu.fig}(a), (b), we present, 
in analogy with Fig.~\ref{Ebbmu.fig}(a), the
results for the vector and scalar toponium soft binding energy,
respectively. The values of the input parameters are the
same as in Figs.~\ref{NmttS1.fig}, \ref{NmttS0.fig},
and in addition $\sigma = 0.33$ according to Eq.~(\ref{signf5}).
The comparative qualitative behavior of the results of
various methods is similar as in the bottonium case,
except that now $R$-PA method appears to fail at
low renormalization scales $\mu \approx m_t \alpha_s \approx 30$ GeV
while $R$-TPS maintains more $\mu$-stability there.

\begin{figure}[htb]
\begin{minipage}[b]{.49\linewidth}
 \centering\epsfig{file=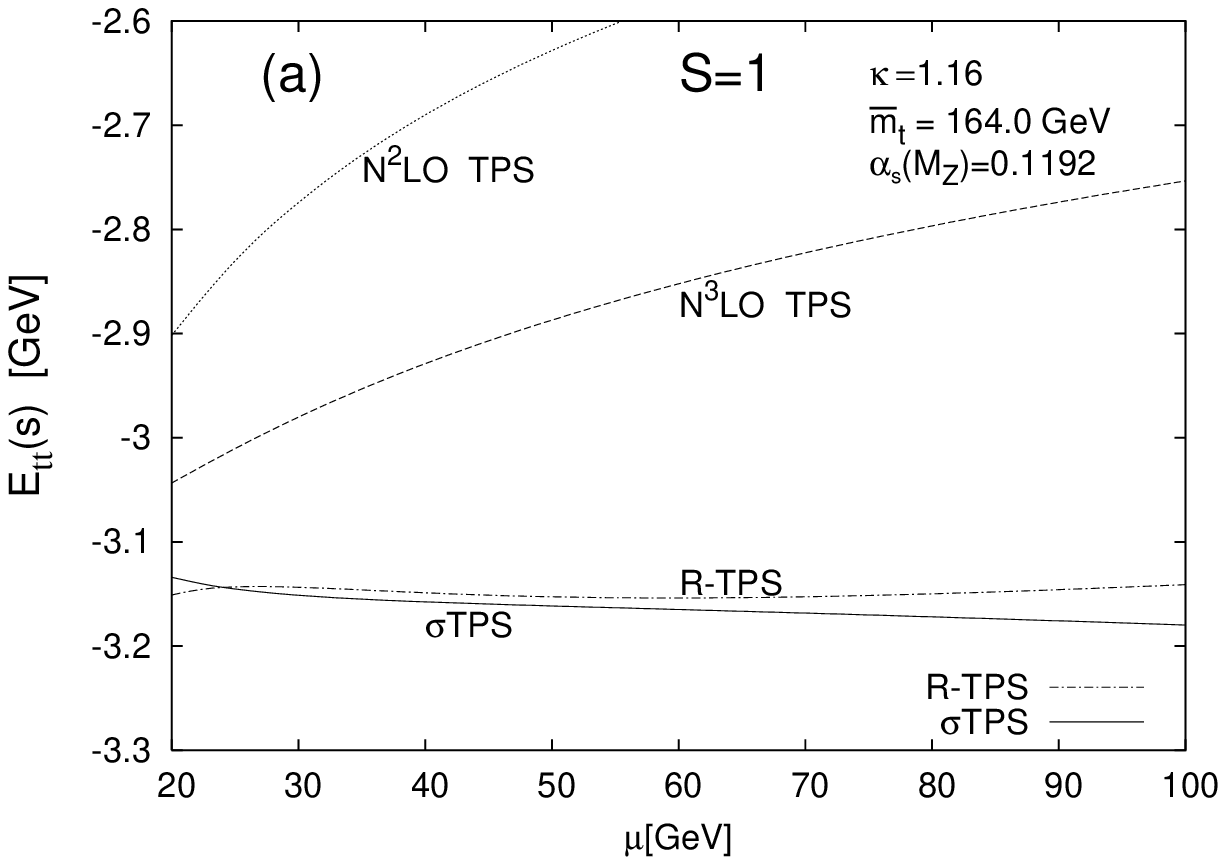,width=\linewidth}
\end{minipage}
\begin{minipage}[b]{.49\linewidth}
 \centering\epsfig{file=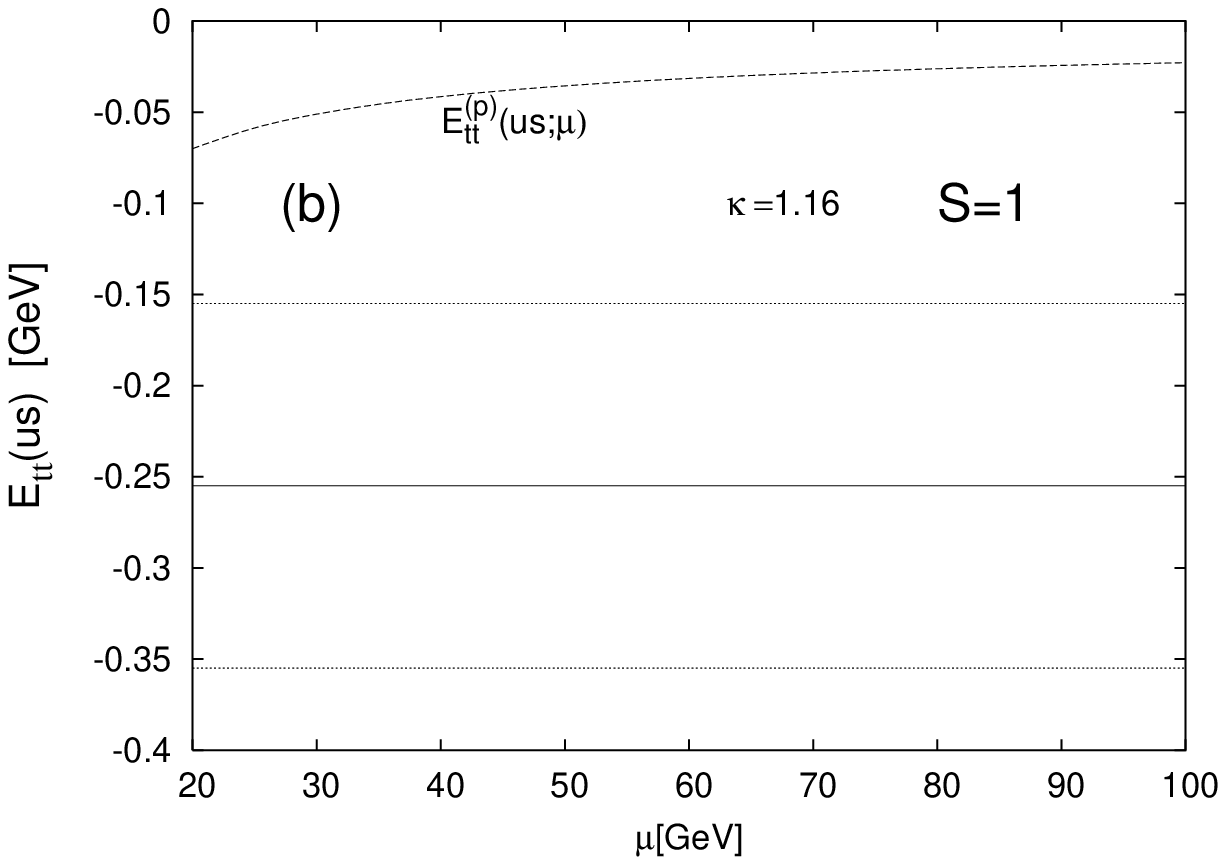,width=\linewidth}
\end{minipage}
\vspace{0.2cm}
\caption{(a) Same as in Fig.~\ref{Ettmu.fig}(a), but 
now the results of the simple TPS evaluation
(\ref{FsTPS}) are included. (b) The ultrasoft
energy parts by different evaluations:
$E_{t \bar t}^{\rm (p)}(us;\mu)$ by Eq.~(\ref{Fusp});
$E_{t \bar t}(us)$ values of Eq.~(\ref{EttusS1}) as
straight lines. The input parameters are the same as in
Fig.~\ref{Ettmu.fig}(a).}
\label{EttTPSmu.fig}
\end{figure}
In Fig.~\ref{EttTPSmu.fig}(a) we present the results
analogous to Fig.~\ref{Ettmu.fig}(a) ($S\!=\!1$ case),
where we now include the results of the
simple TPS evaluation (\ref{FsTPS}) for $t {\bar t}$. 
In Fig.~\ref{EttTPSmu.fig}(b), we present the result for 
for the ``perturbative'' ultrasoft part 
$E_{t \bar t}^{\rm (p)}(us;\mu)$ calculated
according to Eq.~(\ref{Fusp}) for the $t {\bar t}$
system (dashed $\mu$-dependent line).
We further include there the more realistic
estimates obtained
later in Subsection \ref{top} [Eq.~(\ref{EttusS1})].

\subsection{Extraction of bottom mass}
\label{mb}

We need to address now also the problem of evaluating the
ultrasoft part $E_{q \bar q}(us)$ of the ground state
binding energy. The estimate of the perturbative part
is given in Eq.~(\ref{Eqqus2}), where it was
essential to take for the renormalization scale
a $us$ scale $\mu \sim \mu_{us} \sim m_q \alpha_s^2$.

For the bottonium case, this scale is below $1$ GeV,
the energy at which we cannot determine perturbatively
$\alpha_s(\mu)$. This indicates that in the bottonium
the $us$ part of the binding energy has an appreciable
nonperturbative part. The lowest energy at which we can
still determine perturbatively $\alpha_s$ is 
$\mu \approx 1.5$-$2.0$ GeV, 
giving $\alpha_s(\mu) \approx 0.30 - 0.35$.
Although this is a soft scale for $b \bar b$, we will
use this also as an ultrasoft scale. Then by Eq.~(\ref{Eqqus2})
\begin{equation}
E_{b \bar b}(us)^{\rm (p)} \approx 
- \frac{4}{9} {\overline m}_q \pi^2 k_{3,0}(us) a^5(\mu_{us})
\approx (- 150 \pm 100) \ {\rm MeV} \ .
\label{Ebbusp}
\end{equation}
The nonperturbative contribution coming from the
gluonic condensate is given by \cite{Voloshin:hc}
\begin{equation}
E_{b \bar b}(us)^{\rm (np)} \approx 
{\overline m}_b 
\pi^2 \frac{624}{425} 
\left( \frac{4}{3} {\overline m}_b \alpha_s(\mu_{us}) \right)^{-4} 
\langle a(\mu_{us}) G_{\mu \nu} G^{\mu \nu} \rangle
\approx  (50 \pm 35) \ {\rm MeV} \ ,
\label{Ebbusnp}
\end{equation}
where we used ${\overline m}_b = 4.2$ GeV, and the value
of the gluon condensate $\langle (\alpha_s/\pi) G^2 \rangle =
0.009 \pm 0.007 \ {\rm GeV}^4$  \cite{Ioffe:2002be}.
Eqs.~(\ref{Ebbusp}) and (\ref{Ebbusnp}) give
\begin{equation}
E_{b \bar b}(us)^{\rm (p+np)} \approx (-100 \pm 106) \ {\rm MeV} \ ,
\label{Ebbus}
\end{equation}
where the two uncertainties were added in quadrature.
In addition, there are finite charm mass contributions
which have been calculated in Ref.~\cite{Brambilla:2001qk}
(based on the results of 
Refs.~\cite{Gray:1990yh,Eiras:2000rh,Hoang:2000fm}). 
These contributions modify the values of $m_b$ and $E_{b \bar b}$,
resulting in the contribution to the mass
$M_{\Upsilon}(1S) = (2 m_b + E_{b \bar b})$
\begin{equation}
\delta M_{\Upsilon}(1S, m_c \not=0) \approx 25 \pm 10 \ {\rm MeV} \ .
\label{mcnot0}
\end{equation}
The estimates (\ref{Ebbus}), and (\ref{mcnot0})
then give a rough estimate of the $us$ and $m_c\!\not=\!0$ contributions
to the bottonium mass
$\delta M_{\Upsilon}(1S; us+m_c) \approx  (- 75 \pm 106)$ MeV.
The mass of the $\Upsilon(1S)$ vector bottonium ground
state is well measured $M_{\Upsilon}(1S) = 9460$ MeV
with virtually no uncertainty \cite{Hagiwara:fs}.
Therefore, the pure perturbative ``soft'' mass is
\begin{equation}
M_{\Upsilon}(1S; s) = 2 m_b + E_{b \bar b}(s) =  9535 \mp 106 \ {\rm MeV}
\ ,
\label{MUs}
\end{equation}
where the uncertainty $\pm 106$ MeV is the rough estimate
dominated by the uncertainty of the $us$ regime contribution.
Our numerical results for $E_{b \bar b}(s)$ in this Section
and for $m_b$ presented in Sec.~\ref{mass} allow us,
by varying the input value of ${\overline m}_b$, to adjust the
sum $2 m_b\!+\!E_{b \bar b}(s)$ to the value given in
Eq.~(\ref{MUs}). For the soft binding energy
we apply the ``$\sigma$-regularized bilocal methods
$\sigma$-TPS and $\sigma$-PA, and $R$-TPS and $R$-PA,
with the aforementioned ``central'' input parameters:
$\alpha_s(M_Z) = 0.1192$; ${\widetilde \mu} = 1.825$ GeV
($\approx \mu_s$), thus $\alpha_s(\widetilde \mu, n_f\!=\!4) =
0.3263$ [$\alpha_s(\mu_s) = 0.326$]; $N_m = 0.555$; $\kappa = 0.59$;
$\sigma = 0.36$; $a_3/4^3 = 86$; $c_4({\overline {\rm MS}}) = 40$.
For $2 m_b$ we apply the bilocal-TPS and $R$-TPS
method, with renormalization scale $\mu_m/{\overline m}_b = 1$,
both methods giving us very similar results 
[cf.~Fig.~\ref{mqmu.fig} (a)]. 
The bilocal-TPS method is applied for $2 m_b$ when $\sigma$-TPS and
$\sigma$-PA are applied for $E_{b \bar b}(s)$; the $R$-TPS
is applied for $2 m_b$ when $R$-TPS and $R$-PA are applied for
$E_{b \bar b}(s)$ (the same combinations of methods will be
applied in the next Subsec.~\ref{top} to the study of toponium).  
The extracted values
of ${\overline m}_b \equiv 
{\overline m}_b(\mu\!=\!{\overline m}_b)$ are then
\begin{subequations}
\label{mbb}
\begin{eqnarray}
{\overline m}_b & = & 4.225 \pm 0.054 \ {\rm GeV} \qquad 
(\sigma\!-\!\text{TPS}) \ ,
\label{mbbsigTPS}
\\
& = & 4.220 \pm 0.056  \ {\rm GeV} \qquad 
(\sigma\!-\!\text{PA}) \ ,
\label{mbbsigPA}
\\
& = & 4.243 \pm 0.080  \ {\rm GeV} \qquad 
(R\!-\!\text{TPS}) \ ,
\label{mbbRTPS}
\\
& = & 4.235 \pm 0.068  \ {\rm GeV} \qquad 
(R\!-\!\text{PA}) \ .
\label{mbbRPA}
\end{eqnarray}
\end{subequations}
The uncertainties are the combination, in quadrature,
of uncertainties from various sources, shown in
Table \ref{table2} for each of the four methods. 
\begin{table}
\caption{\label{table2} The separate uncertainties 
$\delta {\overline m}_b$ (in MeV) for the extracted
value of ${\overline m}_b$ from various sources:
1.) $us$ [$\delta E_{b \bar b}(us)^{\rm (p+np)} = -100 \pm 106$ MeV];
2.) $\mu = 3 \pm 1$ GeV; 
3.) $\mu_m = {\overline m}_b ( 1 \pm 0.5)$;
4.) $\alpha_s(m_{\tau}) = 0.3254 \pm 0.0125$
[$\alpha_s(M_Z) = 0.1192 \pm 0.0015$];
5.) $N_m = 0.555 \pm 0.020$ [$\kappa = 0.59 \pm 0.19$];
6.) $a_3/4^3 = 86.\pm 23.$;
7.) $c_4 = 40. \pm 60.$;
8. $\sigma = 0.36 \pm 0.03$;
9. $m_c \not= 0$ ($\delta M_{\Upsilon}(m_c\!\not=\!0)=\pm 10$ MeV).}
\begin{ruledtabular}
\begin{tabular}{crrrrrrrrr}
 & $us$ & $\mu$ & $\mu_m$ & $\alpha_s$ & $N_m$ & $a_3$ & $c_4$ & 
$\sigma$ & $m_c$ \\
\hline
$\sigma$-TPS & $- 49$ & $+ 9$ & $- 4$ & $-13$ &$ -3$  &$ + 2$ &
$ - 8$ & $+ 4$ & $- 5$
\\
 & $+ 49$ & $- 13$ & $+ 2$ &$ + 14$ &$ + 2$ &$ - 2$ & $+ 8$ &
$ - 9$ &$ + 5$
\\
\hline
$\sigma$-PA &  $- 49 $ & $ + 13 $ &  $- 4 $ & $ - 15 $ & 
$ -3 $ & $ + 1 $ & $ - 5 $ & $ + 5 $ & $ -  5$ 
\\
 & $ + 49 $ & $ -20 $ & $ + 2 $ & $ + 15 $ & $ + 2 $ & 
 $- 1 $ & $ + 2 $ & $ - 9 $ & $ + 5 $
\\
\hline
$R$-TPS & $ - 50 $ & $ -4 $ & $ + 4 $ & $ - 8 $ &
 $ - 9 $ & $ - 3 $& $0$ & $0$  &  $-5$
\\
 &  $+50 $ & $ + 45 $ & $ - 40 $ & $ + 10 $ &
 $ + 11 $ & $ + 3 $ &$0$ & $0$  &  $+5$
\\
\hline
$R$-PA &  $- 49 $& $ + 3 $ & $ + 4 $ & $ - 11 $ &
 $ -4 $  & $- 2 $  &$0$  &$0$  &  $-5$  
\\
 &  $+ 49 $ & $-20 $ &  $- 40 $ &  $+ 12 $ & 
 $+ 4 $ &  $+ 2 $ &$0$ &$0$ &  $+5$
\\
\end{tabular}
\end{ruledtabular}
\end{table}
In the case of asymetric uncertainties, the larger is taken.
The largest uncertainty ($\pm 0.049$ GeV) comes from the $us$ sector
uncertainty $\pm 0.106$ GeV of Eq.~(\ref{Ebbus}). In the case
of $R$-TPS method, the variation of the 
soft binding energy with the variation of the renormalization
scale is a competing source of uncertainty for
${\overline m}_b$ ($\pm 0.045$ GeV), and in the case of
$R$-TPS and $R$-PA methods (where $m_t$ is resummed by
$R$-TPS) the uncertainty from the variation of the
renormalization scale $\mu_m$ in the $2 m_t$-resummation
is competing as well ($0.040$ GeV). The arithmetic average
of the central values of Eqs.~(\ref{mbb}) gives us
\begin{equation}
{\overline m}_b = 4.231 \pm 0.068 \ {\rm GeV} \qquad
\text{(our average)} \ ,
\label{mbbav}
\end{equation}
where we emphasize that the central value for the
strong coupling parameter was chosen to be
$\alpha_s(M_Z) = 0.1192$. In Eq.~(\ref{mbbav}),
the uncertainty was chosen to be the second largest
uncertainty in Eqs.~(\ref{mbb}). The largest uncertainty,
$\pm 0.080$ GeV of the $R$-TPS method, was discarded
because $R$-TPS is the only one of the four methods
which fails simultaneously at the low $\mu_m$ 
($< {\overline m}_b$) and low $\mu$ ($ < 3$ GeV)
renormalization scales.

\subsection{Numerical results for the toponium}
\label{top}

For the binding energy of the toponium, the numerical results
are obtained in the following way. First the value of the
(PV) pole mass $m_t$ is fixed to the central experimental value 
$m_t = 174.3$ GeV \cite{Hagiwara:fs}. For calculation
of the binding energy, ${\overline m}_t$ is an
input parameter (but not $m_t$).
When $\alpha_s$ varies [$\alpha_s(M_Z) = 0.1192 \pm 0.0015$],
the two methods of Sec.~\ref{mass} [cf.~Fig.~\ref{mqmu.fig} (b)], 
with the renormalization scale $\mu_m = {\overline m}_t$,
give ${\overline m}_t = 164.000^{ - \ 0.153}_{+ \ 0.163}$ GeV
(bilocal method) and 
${\overline m}_t = 164.011^{ - \ 0.153}_{ + \ 0.162}$ GeV 
($R$-method), when $m_t=174.3$ GeV (PV value).
The values of ${\overline m}_t$ change by $0.020$ GeV or less when
the other parameters are varied (renormalization scale $\mu_m$; 
$N_m$ and $c_4$ for bilocal method; see Sec.~\ref{mass}),
and such small variation in ${\overline m}_t$ influences the
toponium binding energy insignificantly\footnote{
A variation $\delta {\overline m}_t \pm 10.0$ MeV results
in $\delta E_{t \bar t}(s) = \mp 0.11$ MeV, when all other input
parameters are kept fixed.}  
-- by less than $0.001$ GeV.

We use as the central ${\overline m}_t$ input value 
${\overline m}_t = 164.000$ GeV
to calculate $E_{t \bar t}(s)$ with the four aforementioned
Borel methods, using for other input parameters their
``central'' values used in Figs.~\ref{Eqqsig.fig}(b), \ref{Ettmu.fig}:
$\alpha_s(M_Z) = 0.1192$; ${\widetilde \mu} = 31$ GeV
($\approx \mu_s$), thus $\alpha_s(\widetilde \mu, n_f=5) =
0.143$ [$\alpha_s(\mu_s) = 0.14$]; $N_m = 0.533$; $\kappa = 1.16 (S\!=\!1),
1.10 (S\!=\!0)$; $\mu = 55$ GeV ($S\!=\!1$), $65$ GeV ($S\!=\!0$);
$\sigma = 0.33$; $a_3/4^3 = 62.5$; $c_4({\overline {\rm MS}}) = 70$.
Then the resulting toponium soft energy is
\begin{subequations}
\label{Etts}
\begin{eqnarray}
E_{t \bar t}(s) & = & - 3.163  \pm 0.116 \ {\rm GeV} \
( - 3.216 \pm 0.120 \ {\rm GeV}) 
\qquad (\sigma\!-\!\text{TPS}) \ ,
\label{EttssigTPS}
\\
& = & - 3.158 \pm 0.115  \ {\rm GeV} \
( - 3.212 \pm 0.118 \ {\rm GeV})
\qquad (\sigma\!-\!\text{PA}) \ ,
\label{EttssigPA}
\\
& = & - 3.154 \pm 0.113  \ {\rm GeV} \
( - 3.200 \pm 0.116 \ {\rm GeV})
\qquad (R\!-\!\text{TPS}) \ ,
\label{EttsRTPS}
\\
& = & - 3.159 \pm 0.115  \ {\rm GeV} \
( - 3.209 \pm 0.118 \ {\rm GeV})
\qquad (R\!-\!\text{PA}) \ ,
\label{EttsRPA}
\end{eqnarray}
\end{subequations}
where the results are given for the vector ($S\!=\!1$) case
and in parentheses for the scalar ($S\!=\!0$) toponium case.
The uncertainties are combinations, in quadrature,
of uncertainties coming from various input sources:
$\delta \alpha_s$, $\delta \mu$, $\delta a_3$,
$\delta c_4$, $\delta N_m$, and $\delta \sigma$.
When $\delta \alpha_s$ is varied, the value
${\overline m}_t$ is varied as well, as described
above, but otherwise it is kept fixed (at $164.000$ GeV).
All the corresponding separate uncertainties
$\delta E_{t \bar t}(s)$ are given in Tables \ref{table3}
for $S\!=\!1$ and \ref{table4} for $S\!=\!0$.
\begin{table}
\caption{\label{table3} The uncertainty 
$\delta E_{t \bar t}(us)$ and the separate uncertainties 
in $\delta E_{t \bar t}(s)$ (in MeV) for the toponium $S\!=\!1$
binding energy from various sources:
1.) $\delta E_{t \bar t}(us)$ [cf.~Eqs.~(\ref{Ettus})];
2.) $\mu = 55 \pm 20$ GeV;
3.) $\alpha_s(M_Z) = 0.1192 \pm 0.0015$;
4.) $N_m = 0.533 \pm 0.020$ 
[$\kappa = 1.16^{\ + \ 0.31}_{\ - \ 0.29}$];
5.) $a_3/4^3 = 62.5 \pm 20.$;
6.) $c_4 = 70. \pm 20.$;
7.) $\sigma = 0.33 \pm 0.03$.
The pole mass $m_t$ is kept fixed at
the value $m_t = 174.30$ GeV.}
\begin{ruledtabular}
\begin{tabular}{crrrrrrr}
 & $\delta E(us)$ & $\delta E(s): \quad \mu$ & $\alpha_s$ & 
$N_m$ & $a_3$ & $c_4$ & 
$\sigma$ \\
\hline
$\sigma$-TPS & $- 100$ & $-7$ & $-105$ & $-21$  &$-4$ &
$ +5$ & $-17$ 
\\
 & $+100 $ & $+8$ & $+109$ & $+23$ & $+4$ & 
$-5$ & $ +29$ 
\\
\hline
$\sigma$-PA &  $-100$ & $-10$ & $-105$ & 
$ -21$ & $-3$ & $+3$ & $-16$ 
\\
 & $+100$ & $+8$ & $+109$ & 
$+22$ & $+3$ & $-2$ & $+26$ 
\\
\hline
$R$-TPS & $-100$ & $+3$ & $-104$ & $ -21$ &
 $ - 8$ &$0$  & $0$ 
\\
 &  $+100$ & $+8$ & $+109$ & $ +26$ &
 $ +8$ & $0$ &$0$ 
\\
\hline
$R$-PA &  $-100$& $ +7$ & $-105$ & $ -18$ &
 $ -6$  &$0$ &$0$ 
\\
 &  $+100$ & $-28$ &  $+110$ &  $+20$ & 
 $+7$ &$0$  &$0$ 
\\
\end{tabular}
\end{ruledtabular}
\end{table}
\begin{table}
\caption{\label{table4} As Table \ref{table3}, but for $S\!=\!0$.
The input parameters are the same, except for
$\mu$ ($= 65 \pm 20$ GeV) and $\kappa$
($= 1.10^{\ + \ 0.39}_{\ - \ 0.33}$, 
corresponding to $N_m = 0.533 \pm 0.020$).}
\begin{ruledtabular}
\begin{tabular}{crrrrrrr}
 & $\delta E(us)$ & $\delta E(s): \quad \mu$ & $\alpha_s$ & 
$N_m$ & $a_3$ & $c_4$ & 
$\sigma$ \\
\hline
$\sigma$-TPS & $- 110$ & $-6$ & $-107$ & $-23$  &$-3$ &
$ +5$ & $-17$ 
\\
 & $+110 $ & $+8$ & $+112$ & $+26$ & $+4$ & 
$-5$ & $ +31$ 
\\
\hline
$\sigma$-PA &  $-110$ & $-8$ & $-107$ & 
$-23$ & $-2$ & $+4$ & $-16$ 
\\
 & $+110$ & $+8$ & $+112$ & 
$+25$ & $+3$ & $-2$ & $+28$ 
\\
\hline
$R$-TPS & $-110$ & $0$ & $-106$ & $ -25$ &
 $ -7$ & $0$ &$0$  
\\
 &  $+110$ & $+13$ & $+111$ & $ +30$ &
 $ +8$ & $0$ &$0$ 
\\
\hline
$R$-PA &  $-110$& $ +9$ & $-107$ & $ -20$ &
 $ -6$  &$0$ &$0$ 
\\
 &  $+110$ & $-27$ &  $+112$ &  $+23$ & 
 $+6$ & $0$ & $0$ 
\\
\end{tabular}
\end{ruledtabular}
\end{table}
The ultrasoft part $E_{t \bar t}(us)$ is
principally perturbative and can be estimated by
formula (\ref{Eqqus2}) where the $us$ coefficient
is given by (\ref{k30us}). This part is more
manageable than in the bottonium case, because
the typical $us$ energy now is still in the perturbative
regime: $\mu_{us} \sim 10^1$ GeV. We determine this
energy by the condition
\begin{equation}
\mu_{us} = \kappa^{\prime} {\overline m}_t \alpha_s^2(\mu_{us}) \ ,
\label{useq}
\end{equation}
where $\kappa^{\prime} \sim 1$. The value
$\kappa^{\prime} = 1$ corresponds to $\mu_{us} \approx 7$ GeV.
Eqs.~(\ref{Eqqus2}) and (\ref{k30us}) then give
for the value $E_{t \bar t}(us) = -0.255$ GeV ($S\!=\!1$)
and $-0.272$ GeV ($S\!=\!0$). When we change to
$\kappa^{\prime} = 2$ ($\mu_{us} = 10.5$ GeV),
the values of $E_{t \bar t}(us)$ go up by $0.100$ and
$0.110$ for the $S\!=\!1,2$, respectively. This
we adopt as the uncertainty in the $us$ sector.
Therefore, we have by Eq.~(\ref{Eqqus2})
\begin{subequations}
\label{Ettus}
\begin{eqnarray}
E_{t \bar t}(us) & = &  - 0.255 \pm 0.100  \ {\rm GeV}
\qquad (S\!=\!1) \ ,
\label{EttusS1}
\\
 & = &  - 0.272 \pm 0.110  \ {\rm GeV}
\qquad (S\!=\!0) \ ,
\label{EttusS0}
\end{eqnarray}
\end{subequations}
corresponding to $\mu_{us} = 7.0^{\ + \ 3.5}_{\ - \ 1.5}$ GeV.
When we take for the soft part $E_{t \bar t}(s)$
the arithmetic average of the results of the four methods
(\ref{Etts}), and combining it with the ultrasoft part
(\ref{Ettus}), we obtain
\begin{subequations}
\label{Ett}
\begin{eqnarray}
E_{t \bar t} & = & - 3.413 \pm 0.153 \ {\rm GeV}
\qquad (S\!=\!1) \ ,
\label{EttS1}
\\
& = &  - 3.481 \pm 0.163  \ {\rm GeV}
\qquad (S\!=\!0) \ .
\label{EttS0}
\end{eqnarray}
\end{subequations}
The two dominant contributions to the uncertainties in Eqs.~(\ref{Ett})
are the uncertainty from $\alpha_s$ in the soft sector,
and the uncertainty of the ultrasoft sector,
as seen from Tables \ref{table3} and \ref{table4} 
and Eqs.~(\ref{Ettus}).

The results (\ref{Ett}) are relevant for the future
determinations of ${\overline m}_t$ from
$t {\bar t}$ production near threshold. We
recall that the determination of the pole mass
$m_t$ has, due to the $b=1/2$ renormalon singularity,
an intrinsic ambiguity of order $\Lambda_{\rm QCD}$,
i.e., several hundred MeV, and cannot be determined from
experiments with a higher accuracy. But the mass
${\overline m}_t$ could be eventually determined with
accuracy of less than 100 MeV, as pointed out
in Refs.~\cite{Kiyo:2000fr} where toponium mass was investigated
using large-$\beta_0$ arguments.
The $S\!=\!1$ toponium state is produced in $e^+ e^-$ annihilation,
while the $S=0$ state in unpolarized $\gamma \gamma$ collisions.
The produced resonance is not exactly at
the ground state mass value $(2 m_t + E_{t \bar t})$ because of the
large decay width of the toponium \cite{Fadin:1987wz,Hoang:2000yr}
\begin{equation}
E_{\rm res.} = 2 m_t + E_{t \bar t} + \delta^{\Gamma} E_{\rm res.} \ .
\label{Eres1}
\end{equation}
The shift in Eq.~(\ref{Eres1}) is
$\delta^{\Gamma} E_{\rm res.} = 100 \pm 10$ MeV
\cite{Penin:2002zv,Penin:1998ik} 
and it is rather stable under the variation of all
input parameters, including $\alpha_s$ and ${\overline m}_t$.
At this point, we should evaluate the sum
$(2 m_t + E_{t \bar t})$ for a general input
value of ${\overline m}_t$ ($\approx 164.$ GeV).
The expected central values of $(2 m_t + E_{t \bar t})$
can be inferred from the
central values of the binding energies (\ref{Ett}) which 
were obtained with the choice ${\overline m}_t = 164.000$ GeV.
We obtain the variation 
\begin{equation}
\delta (2 m_t + E_{t \bar t}) \approx \pm 2.09 \ \delta {\overline m}_t
\ ,
\label{dMdmtb}
\end{equation}
when only the input parameter ${\overline m}_t$ is varied
around its central value $164.00$ GeV, while all the other
input parameters ($\alpha_s$, $N_m$, $\mu_m$, $\mu$, $a_3$, $c_4$,
$\sigma$) are kept fixed at their corresponding central values.\footnote{
More precisely, $\delta {\overline m}_t = \pm 100$ MeV 
would correspond to $\delta (2 m_t + E_{t \bar t}) \approx \pm 208.8$ MeV,
of which $\delta (2 m_t) = \pm 210.1$ MeV,
$\delta E_{t \bar t}(s) = \mp 1.1$ MeV, and
$\delta E_{t \bar t}(us) = \mp 0.2$ MeV.}
At ${\overline m}_t = 164.000$ GeV, the bilocal method
gives $m_t = 174.300$ GeV and the $R$-method
$m_t = 174.288$ GeV. Thus, combining the average of this with
relations (\ref{dMdmtb}) and (\ref{Ett}), we expect the approximate
central values $(2 m_t + E_{t \bar t} ) = 345.175$ GeV
for $S\!=\!1$ and $345.107$ GeV for $S\!=\!0$, when
${\overline m}_t = 164.000$ GeV.
The uncertainties of $(2 m_t + E_{t \bar t} )$
originate from the variation of all the input parameters except 
${\overline m}_t$. Some of them are expected to be close to
the uncertainties in Eqs.~(\ref{Ett}) given for the
binding energies. However, they are not equal to 
these uncertainties of Eqs.~(\ref{Ett}) because the latter 
were obtained by keeping the pole mass fixed ($m_t = 174.3$ GeV).
Now, however, ${\overline m}_t=164.0$ GeV is kept fixed,
and variations of $E_{t \bar t}$ and $m_t$
become correlated in the sum $(2 m_t + E_{t \bar t})$.
More importantly, the variation of $\alpha_s$ now changes
$E_{t \bar t}(s)$ and $2 m_t$, and, to a lesser degree,
$E_{t \bar t}(us)$; the variation of $N_m$ changes $\kappa$
which in turn changes $E_{t \bar t}(us)$ 
[Eqs.~(\ref{Eqqus2}) and (\ref{k30us})] and,
to a lesser degree, $E_{t \bar t}(s)$ and $2 m_t$.
The explicit calculations give for $S\!=\!1$
\begin{subequations}
\label{Mt2S1}
\begin{eqnarray}
(2 m_t + E_{t \bar t} ) &=&  345.181 \pm 0.253 \ {\rm GeV} 
\qquad (\sigma\!-\!\text{TPS} ) \ ,
\label{Mt2S1sigTPS}
\\
& = &  345.186 \pm 0.253 \ {\rm GeV} 
\qquad (\sigma\!-\!\text{PA}) \ ,
\label{Mt2S1sigPA}
\\
& = &  345.168 \pm 0.254 \ {\rm GeV} 
\qquad (R\!-\!\text{TPS}) \ ,
\label{Mt2S1RTPS}
\\
& = &  345.163 \pm 0.256 \ {\rm GeV} 
\qquad (R\!-\!\text{PA}) \ ,
\label{Mt2S1RPA}
\end{eqnarray}
\end{subequations}
and for $S\!=\!0$
\begin{subequations}
\label{Mt2S0}
\begin{eqnarray}
(2 m_t + E_{t \bar t} ) &=&  345.119 \pm 0.263 \ {\rm GeV} 
\qquad (\sigma\!-\!\text{TPS} ) \ ,
\label{Mt2S0sigTPS}
\\
& = &  345.116 \pm 0.263 \ {\rm GeV} 
\qquad (\sigma\!-\!\text{PA}) \ ,
\label{Mt2S0sigPA}
\\
& = &  345.105 \pm 0.261 \ {\rm GeV} 
\qquad (R\!-\!\text{TPS}) \ ,
\label{Mt2S0RTPS}
\\
& = &  345.096 \pm 0.263 \ {\rm GeV} 
\qquad (R\!-\!\text{PA}) \ .
\label{Mt2S0RPA}
\end{eqnarray}
\end{subequations}
Here, the resummation of the mass $2 m_t$ was performed
by the bilocal TPS method in the first two cases
[Eqs.~(\ref{Mt2S1sigTPS}), (\ref{Mt2S1sigPA}) and
(\ref{Mt2S0sigTPS}), (\ref{Mt2S0sigPA})],
and by the $R$-TPS method in the last two cases
[Eqs.~(\ref{Mt2S1RTPS}), (\ref{Mt2S1RPA}) and
(\ref{Mt2S0RTPS}), (\ref{Mt2S0RPA})]
-- cf.~Sec.~\ref{mass}.
In Tables \ref{table5} and \ref{table6} we
give, for $S\!=\!1$ and $S\!=\!0$, respectively, 
separate uncertainties in the mass
$(2 m_t + E_{t \bar t})$ coming from the corresponding
variations of the input parameters
$\alpha_s$, $N_m$, $\mu_m$, $\mu$, $a_3$, $c_4$, $\sigma$
and $\mu_{us}$.
\begin{table}
\caption{\label{table5} The separate uncertainties 
$\delta [2 m_t + E_{t \bar t}(s\!+\!us)]$ (in MeV) for the toponium 
$S\!=\!1$ mass from various sources:
1.) $\mu_{us} = 7.0^{\ - \ 1.5}_{\ + \ 3.5}$ GeV [cf.~Eqs.~(\ref{Ettus})];
2.) $\mu = 55 \pm 20$ GeV;
3.) $\mu_m = {\overline m}_b ( 1 \pm 0.5)$;
4.) $\alpha_s(M_Z) = 0.1192 \pm 0.0015$;
5.) $N_m = 0.533 \pm 0.020$ [$\kappa = 1.16^{\ + \ 0.31}_{\ - \ 0.29}$];
6.) $a_3/4^3 = 62.5 \pm 20.$;
7.) $c_4 = 70. \pm 20.$;
8.) $\sigma = 0.33 \pm 0.03$.
The input mass ${\overline m}_t = 164.00$ GeV is kept fixed.}
\begin{ruledtabular}
\begin{tabular}{crrrrrrrr}
 & $\mu_{us}$ & $\mu$ & $\mu_m$ & $\alpha_s$ & 
$N_m$ & $a_3$ & $c_4$ & $\sigma$ \\
\hline
$\sigma$-TPS & $- 100$ & $-7$ & $+13$ & $+188$ & $+94$  &
$-4$ & $ +3$ & $-17$ 
\\
 & $+100$ & $+8$ & $-9$ & $-203$ & $-108$ & $+4$ & $-3$ &
$ +30$ 
\\
\hline
$\sigma$-PA &  $-100$ & $-10$ & $+13$ & $+189$ & 
$+94$ & $-3$ & $+1$ & $-16$ 
\\
 & $+100$ & $+8$ & $-9$ & $-203$ & $-108$ & 
 $+3$ & $-1$ & $+26$ 
\\
\hline
$R$-TPS & $-100$ & $+2$ & $-9$ & $+188$ & $+54$ &
 $ - 8$ &$0$  & $0$ 
\\
 &  $+100$ & $+7$ & $-95$ & $-203$ & $-65$ &
 $ +8$ & $0$ &$0$ 
\\
\hline
$R$-PA &  $-100$& $ +6$ & $-9$ & $+187$ & $+57$ &
 $-7$  &$0$ &$0$ 
\\
 &  $+100$ & $-29$ & $-95$ & $-202$ &  $-71$ & 
 $+6$ &$0$  &$0$ 
\\
\end{tabular}
\end{ruledtabular}
\end{table}
\begin{table}
\caption{\label{table6} As Table \ref{table5}, but for $S\!=\!0$.
The input parameters are the same, except for
$\mu$ ($= 65 \pm 20$ GeV) and $\kappa$
($= 1.10^{\ + \ 0.39}_{\ - \ 0.33}$, 
corresponding to $N_m = 0.533 \pm 0.020$).}
\begin{ruledtabular}
\begin{tabular}{crrrrrrrr}
 & $us$ & $\mu$ & $\mu_m$ & $\alpha_s$ & $N_m$ & $a_3$ & $c_4$ & 
$\sigma$ \\
\hline
$\sigma$-TPS & $- 110$ & $-7$ & $+13$ & $+184$ & $+112$  &$-4$ &
$ +3$ & $-18$ 
\\
 & $+110$ & $+7$ & $-9$ & $-199$ & $-127$ & $+3$ & $-4$ &
$ +30$ 
\\
\hline
$\sigma$-PA &  $-110$ & $-9$ & $+13$ & $+184$ & 
$+112$ & $-3$ & $+1$ & $-17$ 
\\
 & $+110$ & $+7$ & $-9$ & $-199$ & $-128$ & 
 $+2$ & $-1$ & $+27$ 
\\
\hline
$R$-TPS & $-110$ & $0$ & $-8$ & $+184$ & $+71$ &
 $ -7$ & $0$ &$0$  
\\
 &  $+110$ & $+13$ & $-95$ & $-199$ & $-83$ &
 $ +8$ & $0$ &$0$ 
\\
\hline
$R$-PA &  $-110$& $ +9$ & $-9$ & $+183$ & $+75$ &
 $ -6$  &$0$ &$0$ 
\\
 &  $+110$ & $-27$ & $-95$ &  $-198$ &  $-90$ & 
 $+6$ & $0$ & $0$ 
\\
\end{tabular}
\end{ruledtabular}
\end{table}
Adding them in quadrature, this gave the uncertainties
in Eqs.~(\ref{Mt2S1sigTPS})--(\ref{Mt2S0RPA}).
We take the arithmetic average of the
central values in Eqs.~(\ref{Mt2S1sigTPS})--(\ref{Mt2S1RPA})
for $S\!=\!1$, and of the central values in
Eqs.~(\ref{Mt2S0sigTPS})--(\ref{Mt2S0RPA}) for $S\!=\!0$
\begin{subequations}
\label{Mt3av}
\begin{eqnarray}
(2 m_t + E_{t \bar t} ) &=& 345.175 \pm 0.256 \ {\rm GeV}
\qquad (S\!=\!1) \ ,
\label{Mt3S1av}
\\
(2 m_t + E_{t \bar t} ) &=& 345.109 \pm 0.263 \ {\rm GeV}
\qquad (S\!=\!0) \ .
\label{Mt3S0av}
\end{eqnarray}
\end{subequations}
Combining this with Eq.~(\ref{dMdmtb}) and the aforementioned
shift value $\delta^{\Gamma} E_{\rm res.} = 100 \pm 10$ MeV
in Eq.~(\ref{Eres1}), this gives finally
\begin{subequations}
\label{Eres2}
\begin{eqnarray}
E_{\rm res.} &=& (345.28 \pm 0.26) \ {\rm GeV} + 
2.09 \ ( {\overline m}_t - 164.00 \ {\rm GeV} )
\quad (S=1) \ ,
\label{Eres2S1}
\\
& = & (345.21  \pm 0.26)  \ {\rm GeV} + 
2.09 \ ( {\overline m}_t - 164.00 \ {\rm GeV} )
\quad (S=0) \ ,
\label{Eres2S0}
\end{eqnarray}     
\end{subequations}
In Tables \ref{table5} and \ref{table6} we see that
the major source of uncertainty is from the uncertainty
$\delta \alpha_s(M_Z) = \pm 0.0015$, followed
by the uncertainty of the ultrasoft sector
scale $\delta \mu_{us}$ [cf.~Eqs.~(\ref{Ettus})] and
in the $\sigma$-methods by the uncertainty
in the renormalon residue parameter $\delta N_m = \pm 0.020$
and in $R$-methods
by the uncertainty $\delta \mu_m$ in the renormalization scale
for the resummation of $2 m_t$.

We could adopt in the ultrasoft regime a more conservative
approach, allowing for the parameter $\kappa^{\prime}$
in Eq.~(\ref{useq}) not just to vary from value $1$
up to value $2$, but also to vary down to value $1/2$.
This would correspond to the variation of $\mu_{us}$
from $7$ GeV down to $4.27$ GeV 
[thus increasing $\alpha_s(\mu_{us})$ from $0.198$ to $0.228$,
if keeping $n_f\!=\!5$].
This would increase the uncertainties 
$\pm 0.100$ and $\pm 0.110$ GeV 
in Eqs.~(\ref{Ettus})
to $\pm 0.260$ and $\pm 0.275$ GeV,
respectively. This would give in our results
(\ref{Eres2}) for the $t {\bar t}$ resonance
the increased uncertainties $\ 0.35$ GeV ($S\!=\!1$)
and $\pm 0.36$ GeV ($S\!=\!0$).

The present experimental uncertainty in the pole mass is
$\delta m_t = 5.1$ GeV \cite{Hagiwara:fs},
corresponding to $\delta {\overline m}_t = 4.86$ GeV
(provided we consider $m_t$ to be the Principal Value pole mass).
This implies, according to results (\ref{Eres2}),
the present experimental uncertainty
$(\delta E_{\rm res.})_{\rm exp.} = \pm 10.16$ GeV,
which is still very much above the uncertainties
$\pm 0.26$ GeV (or: $\pm 0.36$ when
conservative approach in the $us$ regime)
coming from the uncertainties of the resummation methods
and of the input parameters (other then ${\overline m}_t$).

In this work we did not include electroweak (Higgs) effects,
which are significant in the case of the top quark.
In Refs.~\cite{Bohm:rj,Jegerlehner:2003py} the 
${\cal O}(\alpha)$ and ${\cal O}(\alpha \alpha_s)$
corrections, respectively, to the relation between $m_t$ and
${\overline m}_t$ mass were calculated.
The size of these corrections significantly depends on the 
hitherto unknown mass $M_H$.
For low Higgs masses $M_H\!=\!100$-$300$ GeV,
these corrections change the value of $m_t$, 
for a given value of ${\overline m}_t$, by several percent. 
Inclusion of these effects would be important for a
realistic extraction of ${\overline m}_t$ from the
resonance energy of the $t \bar t$ production.\footnote{
We thank to M.~Kalmykov for clarifications on this point.}    

\section{Comparisons and conclusions}
\label{summ}

In this Section we will compare our results
with some of the results recently published in the
literature. 

Our results for the mass ${\overline m}_b$,
Eqs.~(\ref{mbb}), (\ref{mbbav}), Table \ref{table2},
will be compared with those recently
obtained by authors who either used pQCD expansions
for the $\Upsilon$(1 S) resonance mass, or $\Upsilon$ sum rules. 
The only input parameter common to all these
methods is $\alpha_s$. The comparison of the various
methods is more reasonable if the same central input
value of (${\overline {\rm MS}}$) $\alpha_s(M_Z)$ is taken. 
Our central value was $\alpha_s(m_{\tau}) = 0.3254$
[$\Rightarrow \alpha_s(M_Z) = 0.1192$] since such \cite{Cvetic:2001ws},
or similar \cite{Geshkenbein:2001mn,Cvetic:2001sn}, 
values follow from the 
(nonstrange) semihadronic $\tau$ decay data which are
very precise \cite{Barate:1998uf}. On the other hand, the 
world average as of September 2002 is
$\alpha_s(M_Z) = 0.1183 \pm 0.0027$ \cite{Bethke:2002rv}.
Most of the authors during the last four years used
central value $\alpha_s(M_Z) \approx 0.118$.
Therefore, for comparisons, we convert our
results (\ref{mbb}) to this central value of $\alpha_s$ --
more specifically, from $\alpha_s(M_Z) = 0.1192 \pm 0.0015$
to $0.1180 \pm 0.0015$. This can be easily done by
inspecting in Table \ref{table2} the column
under $\alpha_s$, giving in Eqs.~(\ref{mbbsigTPS})--(\ref{mbbRPA})
an increase in the central values of $11$, $12$, $8$ and $10$ MeV,
respectively. This gives the average $10$ MeV
higher than in Eq.~(\ref{mbbav})
\begin{equation}
{\overline m}_b = 4.241 \pm 0.068 \ {\rm GeV} \qquad
[ \text{average when:} \  \alpha_s(M_Z) = 0.1180 \pm 0.0015 ] \ .
\label{mbbav2}
\end{equation}
All the separate uncertainties given in Table \ref{table2}
remain, of course, valid also in this translated result.
In Table \ref{table7}, we give comparison of this result
with others in the recent literature. 
\begin{table}
\caption{\label{table7} Recently obtained values of 
(${\overline {\rm MS}}$) ${\overline m}_b$ mass obtained from 
$\Upsilon$ sum rules or from spectrum of the 
$\Upsilon$(1S) resonance. Wherever needed 
(\cite{Penin:2002zv,Lee:2003hh}),
the central mass values were adjusted to the 
common input central value $\alpha_s(M_Z) = 0.118$.}
\begin{ruledtabular}
\begin{tabular}{l c l l}
reference & method & order & ${\overline m}_b({\overline m}_b)$ (GeV)
\\
\hline
PP98 \cite{Penin:1998zh} & $\Upsilon$ sum rules & NNLO &
$4.21 \pm 0.11$
\\
MY98 \cite{Melnikov:1998ug} & $\Upsilon$ sum rules & NNLO &
$4.20 \pm 0.10$
\\
BS99 \cite{Beneke:1999fe}  & $\Upsilon$ sum rules & NNLO &
$4.25 \pm 0.08$
\\
H00 \cite{Hoang:2000fm} & $\Upsilon$ sum rules & NNLO &
$4.17 \pm 0.05$
\\
KS01 \cite{Kuhn:2001dm}  & $\Upsilon$ sum rules & NNLO &
$4.209 \pm 0.050$
\\
CH02 \cite{Corcella:2002uu} & $\Upsilon$ sum rules & NNLO &
$4.20 \pm 0.09$
\\
E02 \cite{Eidemuller:2002wk} &  $\Upsilon$ sum rules & NNLO &
$4.24 \pm 0.10$
\\
P01 \cite{Pineda:2001zq} & spectrum, $\Upsilon$(1S) & NNLO &
$4.210 \pm 0.090 \pm 0.025$
\\
BSV01 \cite{Brambilla:2001qk}  & spectrum, $\Upsilon$(1S) & NNLO &
$4.190 \pm 0.020 \pm 0.025$
\\
PS02 \cite{Penin:2002zv}  & spectrum, $\Upsilon$(1S) &${\rm N}^3{\rm LO}$ &
$4.349 \pm 0.070$
\\
L03 \cite{Lee:2003hh} & spectrum, $\Upsilon$(1S) &${\rm N}^3{\rm LO}$ &
$4.19 \pm 0.04$
\\
this work, Eq.~(\ref{mbbav2})&spectrum, $\Upsilon$(1S)&${\rm N}^3{\rm LO}$&
$4.241 \pm 0.070$ 
\\
\end{tabular}
\end{ruledtabular}
\end{table} 
All these results have
the central value $\alpha_s(M_Z) = 0.118$. Wherever the
central value of $\alpha_s$ was different
\cite{Penin:2002zv,Lee:2003hh}, 
we performed the corresponding translation.
There are two important numerical effects in our result.
The first is the separate evaluation of the ``perturbative'' 
ultrasoft energy part at the corresponding low renormalization energy
($\alt 2$ GeV), Eqs.~(\ref{Eqqus2}) and (\ref{Ebbusp}). 
If we had not separated the (``perturbative'') ultrasoft from
the soft part of the binding energy,
the use of the common renormalization
energy scale $\mu$ ($\approx 3$ GeV) in the resummation then
would have given us the central value of $E_{t \bar t}(us)$
by about $+100$ MeV higher. Then the 
extracted value of ${\overline m}_b$
would have gone down by about $46$ MeV, giving the
value ${\overline m}_b \approx 4.195 \pm 0.068$,
with the central value close to that of L03 in Table \ref{table7}.
On the other hand, that latter value is quite clearly
lower than the value PS02 in Table \ref{table7}, by about $150$ MeV,
principally because of the $b=1/2$ renormalon effect which
were taken into account here and in Ref.~\cite{Lee:2003hh}.
Thus, the renormalon effect brings down the extracted central
value of ${\overline m}_b$ by about 150 MeV, but the separate 
evaluation/estimate of the ultrasoft contribution brings
it up by about $50$ MeV. The renormalon effect can also be
understood from Fig.~\ref{Ebbmu.fig}(b) which suggests that
(at $\mu \approx 3$ GeV) the renormalon effect pushes upward 
the soft binding energy $E_{b \bar b}(s)$ by about $300$ MeV.  
We note that PS02 used pole mass $m_b$ in their 
${\rm N}^3{\rm LO}$ TPS evaluation of
the mass of the $\Upsilon(1S)$ resonance
before extracting the value of ${\overline m}_b$.

Our results for the toponium binding energies are given
in Eqs.~(\ref{Etts}), (\ref{Ettus}) and (\ref{Ett}), 
in connection with Tables \ref{table3} and \ref{table4}.
The result of Ref.~\cite{Lee:2003hh},
was $E_{t \bar t} \approx -3.08 \pm 0.02$ GeV
(for $S\!=\!1$), but the central value of $\alpha_s$
used there was $\alpha_s(M_Z) = 0.1172$.
The result of Ref.~\cite{Penin:2002zv} was
$E_{t \bar t} \approx - 3.01$ GeV, using the
central value $\alpha_s(M_Z) = 0.1185$.
In Table \ref{table8} we present our result (\ref{Ett})
together with the results of these two references,
in both cases rescaled to the common central
$\alpha_s$ value $\alpha_s(M_Z) = 0.1192$.
\begin{table}
\caption{\label{table8} Comparison of some of the toponium
binding energy values $E_{t \bar t}$ recently obtained in the literature.
The first two values were correspondingly rescaled
to our central input $\alpha_s$-value $\alpha_s(M_Z) = 0.1192$,
and $m_t = 174.3$ GeV.}
\begin{ruledtabular}
\begin{tabular}{l c l l}
reference & order & $E_{t \bar t}$ (GeV)
\\
\hline
PS02 \cite{Penin:2002zv}  & ${\rm N}^3{\rm LO}$ &
$-3.065 \pm 0.157$ \ ($S\!=\!1,0$)
\\
L03 \cite{Lee:2003hh} & ${\rm N}^3{\rm LO}$ &
$-3.21 \pm 0.15$ \ ($S\!=\!1$)
\\
this work, Eq.~(\ref{EttS1})& ${\rm N}^3{\rm LO}$&
$-3.413 \pm 0.153$ \ ($S\!=\!1$) 
\\
this work, Eq.~(\ref{EttS0})& ${\rm N}^3{\rm LO}$&
$-3.481 \pm 0.163$ \ ($S\!=\!0$) 
\end{tabular}
\end{ruledtabular}
\end{table}  
We see that in our case the toponium binding energies
are significantly lower. This lowering is 
a combination of the renormalon $b=1/2$ effect
and of the ultrasoft effect. Fig.~\ref{EttTPSmu.fig}(a)
suggests that the renormalon effect, in comparison to
${\rm N}^3{\rm LO}$ TPS, brings down
the soft binding energy $E_{t \bar t}(s)$
by about $150$ MeV ($\mu\!=\!30$ GeV)
and $300$ MeV ($\mu\!=\!60$ GeV).
Further, the ultrasoft effect (\ref{Ettus})
brought down the binding energy by about $200$ MeV.
More specifically, when making our resummation with no separation 
of $s$ and $us$ part, and
using the common renormalization scale $\mu = 50$-$60$ GeV,
would give results for the binding energy $E_{t \bar t}$
higher by about $200$ MeV.
The deviation of our result for $E_{t \bar t}$ from the result of
L03 in Table \ref{table8} can be explained principally with the
ultrasoft effect, and the deviation from PS02 with combination 
of both the ultrasoft and the renormalon effect. 
We note that P02 used in their calculation of $E_{t \bar t}$ 
the ${\rm N}^3{\rm LO}$ TPS with $\mu \approx 30$ GeV and the
pole mass $m_t$.

This lower binding energy $E_{t \bar t}$
is then reflected in the value of the
peak (resonance) position $E_{\rm res.}$ -- Eqs.~(\ref{Eres2})
and Tables \ref{table5} and \ref{table6}. For example,
Ref.~\cite{Penin:2002zv} obtains for $m_t=174.3$ GeV
[and central value $\alpha_s(M_Z)=0.1192$]
the values $E_{\rm res.} = 345.63 \pm 0.16$ GeV
for $S\!=\!1$ and $0$, while we get the values 
$345.28 \pm 0.26$ GeV ($S\!=\!1$) and 
$345.21 \pm 0.26$ GeV ($S\!=\!0$), i.e. lower
by $350$ and $420$ MeV than \cite{Penin:2002zv}.
In Ref.~\cite{Hoang:2000yr}, NNLO results for
$E_{\rm res.}$ of several groups
\cite{Hoang:1999zc,Melnikov:1998pr,Yakovlev:1998ke,Nagano:1999nw,Penin:1998ik}
were compared who used in their calculations various
renormalon-free masses for the top quark.
Their results were taken for
the central input values $\alpha_s(M_Z)=0.1190$ and
${\overline m}_t = 165.00$ GeV,
and are all around $E_{\rm res.} \approx 345.5$ GeV,
with variations, due to the renormalization scale
ambiguity, being usually below $10$ MeV.
For these central input values of $\alpha_s$ and 
${\overline m}_t$, our results (\ref{Eres2}) 
(see also Tables \ref{table5} and \ref{table6}) 
get modified to 
$347.34 \pm 0.26$ GeV ($S\!=\!1$) and
$347.27 \pm 0.26$ GeV ($S\!=\!0$),
i.e., lower by about $200$-$300$ MeV.

\begin{acknowledgments}
We express gratitude to the following persons for helpful communications:
M.~Beneke, V.M.~Braun, M.~Neubert, A.A.~Penin, A.~Pineda, and Y.~Sumino. 
This work was supported in part by FONDECYT (Chile) 
grant No.~1010094 (G.C.), project USM No.~110321 
of the UTFSM (C.C. and G.C), and Mecesup FSM 9901; 
one of us (P.G.) would like
to acknowledge the support of I.~Schmidt.

\end{acknowledgments}

\appendix

\section{Coefficients for the expansion of the soft binding energy}
\label{app:sbe}

We write down here the explicit coefficients $f_j$ of
the expansion (\ref{Fs}) for the soft part of the
ground state binding energy. The logarithms appearing
in these expressions involve three scales 
[$\mu, {\widetilde \mu}, {\overline m}_q$ and 
${\overline \mu}(\widetilde \mu) = 
(4/3) {\overline m}_q \alpha_s(\widetilde \mu)$]
\begin{eqnarray}
L_1 & = & \ln \left( 
\frac{{\overline m}_q}{{\overline \mu}(\widetilde \mu)} \right) \ ,
\qquad
L_2 = \ln \left( \frac{ {\overline m}_q}{\widetilde \mu} \right) \ ,
\qquad
L_{\mu} = \ln \left( \frac{ {\overline m}_q}{\mu} \right) \ .
\label{logs}
\end{eqnarray}
The coefficients $f_j$ are
\begin{eqnarray}
f_1 & = & \frac{1}{2} (35 + 22 L_1 - 11 L_{\mu} - 11 L_2 )
+ \frac{1}{9} (-11 - 6 L_1 + 3 L_{\mu} + 3 L_2) n_f \ .
\label{f1}
\end{eqnarray}
\begin{subequations}
\label{f2}
\begin{eqnarray}
f_2 & = & f_{2}^{(0)} + f_2^{(1)} n_f + f_2^{(2)} n_f^2 \ ,
\label{f2exp}
\\
f_{2}^{(0)}  & = & {\big [}
381.674 + 90.75 L_1^2  + 30.25 L_{\mu}^2  
+ L_1 (246.417 - 121 L_{\mu} - 60.5 L_2) - 48.5 L_2 
\nonumber\\
&&
+ L_{\mu} (-205.25 + 60.5 L_2) - 11.6973 S(S+1) {\big ]} \ ,
\label{f20}
\\
f_{2}^{(1)}  & = & {\big [}
-42.7469 - 11 L_1^2  - 3.66667 L_{\mu}^2  
+ L_{\mu} (26.6944 - 7.33333 L_2) + 6.80556 L_2 
\nonumber\\
&&
+ L_1 (-33.0556 + 14.6667 L_{\mu} + 7.33333 L_2) {\big ]} \ ,
\label{f21}
\\
f_{2}^{(2)}  & = &  {\big [}
1.16286 + (3/9) L_1^2  + (1/9) L_{\mu}^2 
+ L_1 (1 - (4/9) L_{\mu} - (2/9) L_2) 
\nonumber\\
&&
+ L_{\mu} (-0.814815 + (2/9) L_2) - 0.185185 L_2 {\big ]} \ .
\label{f22} 
\end{eqnarray}
\end{subequations}
\begin{subequations}
\label{f3}
\begin{eqnarray}
f_3 & = & f_{3}^{(0)} + f_3^{(1)} n_f + f_3^{(2)} n_f^2 
+ f_3^{(3)} n_f^3 \ ,
\label{f3exp}
\\
f_{3}^{(0)}  & = & {\big [}
6726.11 + 665.5 L_1^3  
- 166.375 L_{\mu} \left( 40.8024 + (-10.5992 + L_{\mu}) L_{\mu} \right)
\nonumber\\ 
&&
+ L_1^2  (2381.5 - 1497.38 L_{\mu} - 499.125 L_2) - 871.429 L_2 
- 499.125 (-1.8843 + L_{\mu}) L_{\mu} L_2  
\nonumber\\ 
&&
- 201.438 L_2^2 + L_1 \left( 7457.17 - 497.292 L_2 + 
L_{\mu} (-4346.38 + 998.25 L_{\mu} + 998.25 L_2) \right) 
\nonumber\\ 
&&
- 257.341 (0.211191 +  L_1 - 0.75 L_{\mu} - 0.25 L_2) S (S+1)
\nonumber\\ 
&&
- 61.4109 \left( -6.13937 + S (S+1) \right) 
\ln \left( \alpha_s(\mu_s) \right) + 440.172 \ln ( \kappa ) 
+ 2 a_3/4^3
{\big ]} \ ,
\label{f30}
\\
f_{3}^{(1)}  & = & {\big [}
-1274.33 - 1277.92 L_1 - 471.125 L_1^2  - 121 L_1^3  + 1182.32 L_{\mu} 
+ 843.667 L_1 L_{\mu} + 272.25 L_1^2  L_{\mu} 
\nonumber\\ 
&&
- 335.813 L_{\mu}^2  - 181.5 L_1 L_{\mu}^2  
+ 30.25 L_{\mu}^3  + 124.501 L_2 + 108.361 L_1 L_2 + 90.75 L_1^2  L_2 
\nonumber\\ 
&&
-186.708 L_{\mu} L_2
- 181.5 L_1 L_{\mu} L_2 + 90.75 L_{\mu}^2  L_2 + 36.7292 L_2^2  
\nonumber\\ 
&&
+ (4.06858 + 15.5964 L_1  - 11.6973 L_{\mu}  - 3.8991 L_2) S (S+1) 
{\big ]} \ ,
\label{f31}
\\
f_{3}^{(2)}  & = & {\big [}
70.8992 + 70.2453 L_1 + 28.9722 L_1^2  + 7.33333 L_1^3  - 65.9925 L_{\mu} 
-51.6667 L_1 L_{\mu} 
\nonumber\\ 
&&
- 16.5 L_1^2  L_{\mu} + 20.5972 L_{\mu}^2 + 11  L_1 L_{\mu}^2  
-1.83333 L_{\mu}^3  - 5.19388 L_2 - 6.57407 L_1 L_2 
\nonumber\\ 
&&
- 5.5 L_1^2  L_2 + 10.9167 L_{\mu} L_2 + 11  L_1 L_{\mu} L_2 
- 5.5 L_{\mu}^2  L_2 - 2.09722 L_2^2
{\big ]} \ ,
\label{f32}
\\
f_{3}^{(3)}  & = & {\big [}
-1.21475 - 1.21714 L_1 - (5/9) L_1^2  - 0.148148 L_1^3  
+  1.16286 L_{\mu} + L_1 L_{\mu} 
+ (1/3) L_1^2  L_{\mu} 
\nonumber\\ 
&&
- 0.407407 L_{\mu}^2
- (2/9) L_1 L_{\mu}^2  + 0.037037 L_{\mu}^3 + 0.0542857 L_2 
+ (1/9) L_1 L_2 + (1/9) L_1^2  L_2 
\nonumber\\ 
&&
- 0.185185 L_{\mu} L_2 - (2/9) L_1 L_{\mu} L_2 
+ (1/9) L_{\mu}^2  L_2 + 0.037037 L_2^2
 {\big ]} \ .
\label{f33}
\end{eqnarray}
\end{subequations}

\end{document}